%% file: draft.tex
\documentclass[10pt,sigconf]{acmart}

\copyrightyear{2020}
\acmYear{2020}
\usepackage{graphicx}
\usepackage{balance}  

\usepackage{url}

\usepackage[titlenumbered,ruled,noend,linesnumbered]{algorithm2e}
\usepackage{amsmath}
\usepackage{amsthm}
\usepackage{amssymb}
\usepackage{comment}
\usepackage{multirow}
\usepackage{hhline}
\usepackage{color}

\usepackage{enumitem}

\DontPrintSemicolon

\def\script#1{\mathcal{#1}}

\def\mS{\script{S}}

{\hspace*{\fill}$\Box$\par}

\newcommand{\squishlisttwo}{
   \begin{list}{$\bullet$}
       { \setlength{\itemsep}{0pt}      \setlength{\parsep}{3pt}
       \setlength{\topsep}{3pt}       \setlength{\partopsep}{0pt}
      \setlength{\leftmargin}{1.5em} \setlength{\labelwidth}{1em}
       \setlength{\labelsep}{0.5em} } }

\newcounter{Lcount}
\newcommand{\squishlist}{
\begin{list}{\arabic{Lcount}. }
{ \usecounter{Lcount}
\setlength{\itemsep}{0pt}
\setlength{\parsep}{0pt}
\setlength{\topsep}{0pt}
\setlength{\partopsep}{0pt}
\setlength{\leftmargin}{2em}
\setlength{\labelwidth}{1.5em}
\setlength{\labelsep}{0.5em} } }

\newcommand{\squishend}{
\end{list} }

\begin{document}
	
	\fancyhead{}

\newtheorem{problem}[theorem]{Problem}
\newtheorem{principle}[theorem]{Principle}
\newtheorem{assumption}[theorem]{Assumption}
\newtheorem{remark}[theorem]{Remark}
\newtheorem{property}[theorem]{Property}

\title{Usable \& Scalable Learning Over Relational Data With Automatic Language Bias}

\author{Jose Picado}
\affiliation{%
	\institution{Oregon State University}
}
\email{picadolj@oregonstate.edu}

\author{Arash Termehchy}
\affiliation{%
	\institution{Oregon State University}
}
\email{termehca@oregonstate.edu}

\author{Sudhanshu Pathak}
\affiliation{%
	\institution{Oregon State University}
}
\email{pathaks@oregonstate.edu}

\author{Alan Fern}
\affiliation{%
	\institution{Oregon State University}
}
\email{alan.fern@oregonstate.edu}

\author{Praveen Ilango}
\affiliation{%
	\institution{Oregon State University}
}
\email{ilangop@oregonstate.edu}

\author{Yunqiao Cai}
\affiliation{%
	\institution{Oregon State University}
}
\email{caiy@oregonstate.edu}

\begin{abstract}

	Relational databases are valuable resources for learning novel and interesting relations and concepts. 
	In order to constraint the search through the large space of candidate definitions, users must tune the algorithm by specifying a language bias. Unfortunately, specifying the language bias is done via trial and error and is guided by the expert's intuitions. 
	We propose AutoBias, a system that leverages information in the schema and content of the database to automatically induce the language bias used by popular relational learning systems. 
	We show that AutoBias delivers the same accuracy as using manually-written language bias by imposing only a slight overhead on the running time of the learning algorithm. 
\end{abstract}


\maketitle

	\input{1.Introduction}

	\input{2.Background}

	\input{3.LearningAlgorithm}

	\input{4.System}

	\input{5.Sampling}

	\input{6.Experiments}
	\input{7.Related}

	\input{7.Conclusion}
\bibliographystyle{ACM-Reference-Format}
\bibliography{../ref}
\end{document}

%% file: 1.Introduction.tex
\section{Introduction}
\label{section:automode-introduction}

Learning novel concepts or relations over relational databases 
has attracted a great deal of attention ~\cite{DeRaedt:2010:LRL:1952055,MLBase:CIDR,Kumar:2015:LGL:2723372.2723713,QuickFOIL}.
Consider the UW-CSE database ({\it alchemy.cs.washington.edu/\\data/uw-cse}), which contains information about a computer science department and its schema fragments are shown in Table~\ref{table:automode-uwcse}.
One may want to predict
the new relation {\it advisedBy(stud,prof)}, which 
indicates that the student {\it stud} is advised by professor {\it prof}. 
Given the UW-CSE database and positive and negative training examples of the {\it advisedBy} relation, relational learning algorithms attempt to find a definition of this relation in terms of the existing relations in the database~\cite{StarAIW,SRLNIPS,lao-etal-2015-learning,10.1145/3183713.3199515,DeRaedt:2010:LRL:1952055,castor:SIGMOD17}. 
Learned definitions are usually first-order logic formulas and often restricted to Datalog programs.
For example, a relational learning algorithm may learn the following Datalog program for the {\it advisedBy} relation:
\begin{align*}
\mathit{advisedBy}(x,y) \leftarrow  \mathit{publication}(z,x),  \mathit{publication}(z,y)
\end{align*}
which indicates that a student is advised by a professor if they have been co-authors of a publication.

Relational learning algorithms can exploit the relational structure of the data, making them useful for domains where structure of data is important
\cite{StarAIW,SRLNIPS,lao-etal-2015-learning,10.1145/3183713.3199515}. 
First, other learning methods, such as logistic regression, rely on the assumption that
the underlying data has IID property, i.e., the data points are independent and taken from the same identical distribution \cite{MachineLearning:Mitchell}.
It is well established that IID assumption is usually violated over relational data, therefore, 
using usual these methods may result in biased models with low testing accuracy over relational data \cite{10.1145/3183713.3199515,SRLNIPS,DeRaedt:2010:LRL:1952055,Getoor:SRLBook}. 
Using these methods may not differentiate between outliers and important relationships between 
different entities in the domain \cite{10.5555/1625275.1625397}. 
Second, their learned definitions are interpretable and easy to understand. 
Third, as they directly leverage the structure of the data, users do not need to perform lengthy and cumbersome 
process of feature engineering. Since methods, such as logistic regression, are designed for the cases where the data is stored
in a single table, their effectiveness rely heavily on the skills of the feature engineers on converting and 
integrating relevant pieces of information from multiple relations in the schema in a single table
\cite{DeRaedt:2010:LRL:1952055,Getoor:SRLBook}. If the engineers miss the important information or 
to aggregate them in a wrong way, the learning method will deliver inaccurate results. 
Relational learning methods are also used to learn features over relational data for downstream non-relational learning methods \cite{lao-etal-2015-learning}. 
Thus, they have been widely used to learn over relational data with applications to designing usable query interfaces \cite{Abouzied:PODS:13,Maier:VLDB:2015,Kalashnikov:2018:FFQ:3183713.3183727}, information extraction \cite{StarAIW,10.1145/3183713.3199515}, 
and entity resolution \cite{Evans2018LearningER}.

The space of possible hypotheses that a relational learning algorithm can explore consists of all Datalog programs defined over the schema of the input database.
This space can be very large if the schema of the input database contains many relations or many attributes.
Therefore, users must constraint the hypothesis space of relational learning algorithms using a {\it language bias}.
One form of language bias is {\it syntactic bias}, which restricts the structure and syntax of the learned Datalog programs.
Relational learning systems usually allow users to specify the syntactic bias through statements called {\it predicate definitions} and {\it mode definitions}~\cite{DeRaedt:2010:LRL:1952055}. 
Predicate and mode definitions express several types of restrictions on the structure of the learned Datalog programs.
Consider the UW-CSE database ({\it alchemy.cs.washington.edu/data/uw-cse}), which contains information about a computer science department and whose schema is shown in Table~\ref{table:automode-uwcse}.
 Table~\ref{table:uwcse-modes} shows a fragment of predicate and mode definitions used for the UW-CSE database. 
Intuitively, predicate definitions restrict the relations that can join in the learned Datalog program and under which attributes. 
For instance, it makes sense to join relations {\it student} and {\it inPhase} under attributes {\it student[stud]} and {\it inPhase[stud]}, but it does not makes sense to join these relations under attributes {\it student[stud]} and {\it inPhase[phase]}. Therefore attributes {\it student[stud]} and {\it inPhase[stud]} are assigned the same type \texttt{T1} and attribute {\it inPhase[phase]} is assigned  a different type \texttt{T3}.
Mode definitions restrict the join paths that can be explored by the Datalog programs and whether attributes can appear as variables or constants.
Relational learning algorithms use mode definitions to restrict the Datalog programs that are explored.
A detailed explanation of predicate and mode definitions is given in Section~\ref{section:automode-languagebias}.
To the best of our knowledge, all (statistical) relational learning systems require some form of syntactic bias to restrict the hypothesis space.

\begin{table}
	\centering
	\caption{ Schema for the UW-CSE dataset. }
\vspace{-10pt}
	\begin{tabular} { l l}
		\hline
		student(stud) & professor(prof) \\
		inPhase(stud, phase) & hasPosition(prof, position)\\
		yearsInProgram(stud, years) &  taughtBy(course, prof, term)\\
		courseLevel(course, level) & ta(course, stud, term) \\
		publication(title, person) & \\
		\hline
	\end{tabular}
	\label{table:automode-uwcse}
\end{table}

\begin{table}
	\centering
	\caption{ A subset of predicate and mode definitions for the UW-CSE dataset. }
	\vspace{-10pt}
	\begin{tabular} { l l }
		\hline
		Predicate definitions & Mode definitions \\
		\hline
		student(T1) & student(+) \\
		inPhase(T1,T2) & inPhase(+,-) \\ 
		professor(T3) & inPhase(+,\#) \\
		hasPosition(T3,T4) & professor(+) \\ 
		publication(T5,T1) & hasPosition(+,-) \\
		publication(T5,T3) & publication(-,+) \\
		\hline
	\end{tabular}
	\label{table:uwcse-modes}
\end{table}

For a relational learning algorithm to be effective and efficient, predicate and mode definitions must encode a great deal of information about the structure of the learned Datalog programs \cite{DeRaedt:2010:LRL:1952055}.
A user should both know the internals of the learning algorithm and the schema of the input database and have a relatively clear intuition on the structure of effective Datalog programs for the target relation to set a sufficient degree of restriction. 
However, there may not be any user that both knows the database concepts, such as schema, and has a clear intuition about the target relation. 
Furthermore, 
the number of predicate and mode definitions of is generally large and hard to debug and maintain.
Users normally improve the initial set of definitions via trial and error, which is a tedious and time-consuming process.
Hence, it takes a lot of time and effort to write and maintain these definitions, particularly for a relatively complex schema.
In our conversations with (statistical) relational learning experts, they have called predicate and mode definitions the ``black magic'' needed to make relational learning work and believe them to be a major reason for the difficulty of working with these systems and their relative unpopularity among users. 

In this paper, we propose a novel approach that leverages the information in the schema and content of the database to generate predicate and mode definitions automatically. Our method uses the exact and approximate database constraints and dependencies to find promising 
patterns in the data. These constraints are usually available in the schema of the database. They can also be discovered from the database instance  
if they are {\it not} stored in the database schema \cite{Papenbrock:2015:DCI:2752939.2752946,Abedjan:2015:PRD:2811716.2811766}. 
We show that the predicate and mode definitions produced by our method deliver the same accuracy as the manually written and tuned ones by experts. 

The automatically induced predicate and mode definitions may not limit the space of the search for the learning algorithm as tightly as the ones written manually by the experts and may result in an under-restricted hypothesis space.
Therefore, using our automatically generated predicate and mode definitions, it may be extremely time-consuming to learn over large databases.
To address this issue, we investigate sampling techniques over the hypothesis space 
of the relational learning algorithms to learn accurate definitions efficiently. 
Currently, relational learning systems sample hypotheses, i.e., Datalog clauses, 
from the underlying database without 
considering their relationships and properties to deal with huge hypothesis space.
We show that this approach results in sampled clauses that are neither a random sample of the relational database nor represent the diversity of the patterns within the data. 
We propose novel sampling methods to address these issues.
We show that our methods deliver considerably more effective results in significantly faster time than the current sampling method used by relational learning algorithms over large databases.
More specifically, our contributions in this paper are as follows.
\begin{itemize}
\item We introduce the problem of language bias for relational learning automatically.

\item We propose a new system called {\it AutoBias}, which leverages the information in the schema and content of the underlying databases to induce the language bias automatically (Section~\ref{section:automode-system}).
AutoBias leverages the exact and approximate inclusion dependencies, i.e., referential integrities, \cite{AliceBook}, in a database to induce predicate definitions for learning concepts over the database (Section~\ref{section:automode-system}).
AutoBias also uses the information in the content of the database to generate mode definitions (Section~\ref{section:automode-predicate-definitions}).

\item To scale AutoBias for large databases, we propose a sampling method that leverages random sampling techniques to construct clauses that connect multiple relations and produce representative patterns from the underlying database efficiently (Section~\ref{section:automode-random-sampling}). 
Since the randomly sampled clauses may be biased toward more connected relations in the database,
it may reduce the effectiveness of relational learning for large and diverse datasets. 
We propose a stratified sampling method to ensure that the 
sampled clause is not biased to some specific tables or patterns and a fair representative of the relevant information in the database (Section~\ref{section:automode-stratified-sampling}). 
We also investigate using these techniques in evaluating the quality of a hypothesis, i.e., 
whether a hypothesis covers sufficiently many positive and few negative examples, during learning
efficiently.


\item We empirically evaluate our language bias induction and sampling techniques over real-world and large databases.
Our empirical study indicates that our proposed language bias generation method delivers almost as accurate results as the ones developed by experts over multiple datasets.
They also show that random sampling approach improves the efficiency of our system significantly and delivers more effective or as effective results than the state-of-the-art sampling techniques over large databases.
It also indicate that stratified sampling delivers a more effective result that that of random sampling 
if the dataset is large and the target relation is complex and is captured by a diverse set of clauses.

\end{itemize}


%% file: 2.Background.tex
\section{Background}
\label{section:automode-background}

\subsection{Basic Definitions}
An {\it atom} is a formula in the form of 
$R(e_1, \ldots, e_n)$,
where $R$ is a relation symbol.
A {\it literal} is an atom, or the negation of an atom.
Each attribute in a literal is set to either a variable or a constant, i.e., value.
Variable and constants are also called {\it terms}.
A {\it Horn clause} (clause for short) is a finite set of literals that contains exactly one positive literal
called {\it head-literal}. Horn clauses are also called conjunctive queries.
A {\it Horn definition} is a set of Horn clauses with the same head-literal.

A relational learning algorithm learns 
a Horn definition from input relational databases and training data. 
The learned definition is called the hypothesis, which is usually restricted to non-recursive Datalog definitions 
without negation, i.e., unions of conjunctive queries, 
for efficiency reasons. 
The {\it hypothesis space} is the set of all candidate Horn definitions that the algorithm can explore.
Each member of the hypothesis space is a {\it hypothesis}.
Given a database instance $I$, clause $C$ {\it covers} example $e$ if $I \wedge C \models e$, where $\models$ is the entailment operator, i.e., if $I$ and $C$ are true, then $e$ is true. 
Definition $H$ covers an example $e$ if at least one its clauses covers $e$.
Relational learning algorithms search over the hypothesis space to find a definition that covers as many positive examples as possible, while covering the fewest possible negative examples.

\subsection{Language Bias}
\label{section:automode-languagebias}

In relational learning algorithms, language bias restricts the structure and syntax of the generated clauses.
Language bias is specified through predicate and mode definitions~\cite{DeRaedt:2010:LRL:1952055}.

\subsubsection{Predicate Definitions}
Predicate definitions assign one or more {\it types} to each attribute in a database relation. 
In a candidate clause, two relations can be joined over two attributes (i.e., attributes are assigned the same variable) only if the attributes have the same type. 
For instance, in Table~\ref{table:uwcse-modes}, the predicate definition \texttt{student(T1)} indicates that the attribute in relation {\it student} is of type \texttt{T1}, and the predicate definition \texttt{inPhase(T1,T2)} indicates that the first and second attributes of relation {\it inPhase} are of type \texttt{T1} and \texttt{T2}, respectively.
Therefore, relations {\it student} and {\it inPhase} can be joined on attributes {\it student[stud]} and {\it inPhase[stud]}.
It is possible to assign multiple types to an attribute. 
For example the predicate definitions 
\texttt{publication(T5,T1)} and \texttt{publication(T5,T3)} indicate that the attribute 
{\it author} in relation {\it publication} belongs to both 
types \texttt{T1} and \texttt{T3}.
Predicate definitions restrict the joins that can appear in a candidate clause:
two relations can be joined only if their attributes share a type.

Intuitively, predicate definitions should assign the same types to attributes that refer to entities of the same {\it semantic type}. 
For instance, attributes {\it student[stud]} and {\it inPhase[stud]} both refer to the entity type {\it student}. Therefore, predicate definitions should assign the same type to these attributes.
On the other hand, attribute {\it inPhase[phase]} refers to entities of type {\it phase}. Therefore, this attribute should be of a different type.
Note that relying on attribute names would not be a reliable way to inferring the semantic types of entities stored in an attribute.
A user should know the schema of the database and the meaning of all attributes in order to write effective predicate definitions.

\subsubsection{Mode Definitions}
Mode definitions indicate whether a term in an literal should be a new variable, i.e., existentially quantified variable, an existing variable, i.e., appears in a previously added literal, or a constant.
They do so by assigning one or more symbols to each attribute in a relation. 
{\it Symbol $+$} indicates that a term must be an existing variable.
{\it Symbol $-$} indicates that a term can be an existing variable or a new variable.
For instance, the mode definition \texttt{inPhase(+,-)} in Table~\ref{table:uwcse-modes} indicates that the first term must be an existing variable and the second term can be either an existing or a new variable.
{\it Symbol $\#$} indicates that a term should be a constant.
For instance, the mode definition \texttt{inPhase(+,\#)} indicates that the second term must be a constant.

Mode definitions restrict the candidate clauses that are explored by the learning algorithm. 
Each literal in a candidate clause must satisfy at least one mode definition.
Some mode definitions do not add any value to the creation of candidate clauses.
For instance, mode definition \texttt{inPhase(+,+)} means that both variables in a new literal must be existing variables. The same literal can be created from mode definitions \texttt{inPhase(+,-)} or \texttt{inPhase(-,+)}. Therefore, mode definition \texttt{inPhase(+,+)} does not add new more information to the candidate clause.
On the other hand, mode definition \texttt{inPhase(-,-)} means that both variables in a literal must be new variables. In this case, the new literal would not be connected to any previously added literal, resulting in a Cartesian product in the clause.
A user should know the learning algorithm and have an intuition of the desired hypotheses in order to write effective mode definitions.

We explain how predicate and mode definitions are used in the learning algorithm in Section~\ref{section:cator-bc}.


%% file: 3.LearningAlgorithm.tex
\section{AutoBias Learning Algorithm}

AutoBias uses the same learning algorithm as existing relational learning algorithms~\cite{progolem,castor:SIGMOD17}.
In this section, we explain this algorithm and how it uses language bias to learn efficiently.
Similar to other relational learning algorithms, AutoBias is a sequential covering algorithm~\cite{ILP20:2012,Quinlan:FOIL,progol,progolem,QuickFOIL,castor:SIGMOD17}.
Algorithm \ref{algorithm:castor-covering} depicts a general sequential covering algorithm.   
The algorithm constructs one clause at a time using the {\it LearnClause} function.
If the clause satisfies the minimum criterion, 
it adds the clause to the learned definition 
and discards the positive examples covered by the definition.
It stops when all positive examples are covered by the learned definition.

\begin{algorithm}
	\SetKwInOut{Input}{Input}
	\SetKwInOut{Output}{Output}
	\Input{Database instance $I$, positive examples $E^+$, negative examples $E^-$}
	\Output{A Horn definition $H$}
	$H = \{\}$\;
	$U = E^+$\;
	\While{$U$ is not empty }{
		$C = LearnClause(I,U,E^-)$\;
		\If{$C$ satisfies minimum criterion} {
			$H = H \cup C$\;
			$U = U - \{ e \in U | H \wedge I \models e \}$\;
		}
	}
	return $H$ \;
	\caption{Sequential covering algorithm.}
	\label{algorithm:castor-covering}
\end{algorithm}

There are two main approaches in developing the {\it LearnClause} function.
In the {\it top-down} approach, the algorithm starts with an empty definition and iteratively add literals to the definition until the definition cannot be improved~\cite{ILP20:2012,Quinlan:FOIL,QuickFOIL}. 
In the {bottom-up} approach, the algorithm first finds 
relevant patterns in the data and then generalizes them to find clauses that capture the training examples
accurately \cite{ILP20:2012,golem,progol,progolem,castor:SIGMOD17}. 
AutoBias follows the latter approach.
Under this approach, the {\it LearnClause} function contains two main steps. 
In the first step, it constructs the most specific clause that covers a positive example, relative to the database.
This clause is called the {\it bottom-clause}.
In the second step, it generalizes the bottom-clause to cover more positive examples, while covering the fewest possible negative examples.
It is shown that the bottom-up approach usually delivers more effective results than those of top-down methods
\cite{progol,progolem,DeRaedt:2010:LRL:1952055,castor:SIGMOD17}.
We now explain each step in more detail.


\subsection{Bottom-clause Construction}
\label{section:cator-bc}


A {\it bottom-clause} $C_e$ associated with an example $e$ is the most specific clause in the hypothesis space that covers $e$ relative to the underlying database $I$. 
The bottom-clause construction algorithm consists of two phases.
1) Find all the information in $I$ relevant to $e$.
The information relevant to example $e$ is the set of tuples $I_e \subseteq I$ that are connected to $e$.
2) Given $I_e$, create the bottom-clause $C_e$.
The bottom-clause construction algorithm is depicted in Algorithm~\ref{algorithm:bottom-clause-original}.
Relational learning algorithms use predicate and mode definitions to restrict the structure and syntax of the bottom-clauses~\cite{Quinlan:FOIL,progol,progolem,castor:SIGMOD17}. 
We now explain the bottom-clause construction algorithm and how it uses predicate and mode definitions.


Assume that we want to create the bottom-clause for example $\mathit{e}$, relative to database $\mathit{I}$.
The algorithm maintains a hash table that maps constants to variables.
The algorithm first assigns new variables to constants in example $\mathit{e}$, 
and inserts the mapping from constants to variables in the hash table.
It creates the head of the bottom-clause by replacing the constants in $\mathit{e}$ with their assigned variables.
Then, for each constant $a$ in the hash table, the algorithm looks for relations that contain attributes with the same type as $a$ and then searches for tuples in these relations that contain constant $a$.
The type of a constant is determined by the attribute in which the constant appears. The attribute types are assigned using {\bf predicate definitions}.
To further restrict the search, the algorithm only considers attributes which contain symbol $+$, according to the {\bf mode definitions}.
For each tuple, the algorithm creates one or more literals 
with the same relation name as the tuple and 
adds the literals to the body of the bottom-clause.
The algorithm also uses mode definitions to determine whether an attribute in a literal should be a variable or a constant.
An attribute $A$ in relation $R$ can be a variable if the mode definitions for relation $R$ contain symbols $+$ or $-$ on attribute $R$.
Attribute $A$ can be a constant if the mode definitions for relation $R$ contain symbols $\#$ on attribute $R$.
If an attribute $A$ can be both a variable and a constant, the algorithm creates two new literals, one literal containing a variable in attribute $A$ and the containing a constant in attribute $A$.
If an attribute should be a variable according to mode definitions, and the constant in this attribute is new, 
the algorithm assigns a new variable to the constant
and adds the new mapping to the hash table.
In the following iterations, the algorithm selects 
tuples in the database that contain the newly 
added constants to the hash table and adds their 
corresponding literals to the clause.
The algorithm finishes after the user-specified number of iterations.

\begin{algorithm}
	\SetKwInOut{Input}{Input}
	\SetKwInOut{Output}{Output}
	\Input{example $e$, \# of iterations $d$, sample size $s$}
	\Output{bottom-clause $C_e$}
	
	$I_{e} = \{\}$\;
	$M = \{\}$ // $M$ stores known constants\;
	
	add constants in $e$ to $M$\;
	
	\For{$i=1$ to $d$}{
		\ForEach{relation $R \in I$}{
			\ForEach{attribute $A$ in $R$}{
				$I_R = \sigma_{A \in M}(R)$\;
				
				\ForEach{tuple $t \in I_R$}{
					add $t$ to $I_{e}$ and constants in $t$ to $M$\;
				}
			}
		}
	}
	$C_e=$ create clause from $e$ and $I_{e}$\;
	return $C_e$\;
	
	\caption{Bottom-clause construction.}
	\label{algorithm:bottom-clause-original}
\end{algorithm}

\begin{example}
	\label{example:bottomclause}
	Consider the database instance $I$ in Table~\ref{table:automode-example-database}, the predicate and mode definitions in Table~\ref{table:uwcse-modes}, and a positive example $e=$ {\it advisedBy(alice,bob)}. 
	Given that the user sets the number of iterations $d$ to 1, the bottom-clause associated with $e$ and relative to $I$ is:
	\begin{align*}
	\mathit{ad}&\mathit{visedBy}(x,y) \leftarrow  \mathit{student}(x), \mathit{professor}(y), \\
	& \mathit{inPhase}(x,u), \mathit{inPhase}(x,\mathit{post\_quals}),  \mathit{hasPosition}(y,v), \\
	& \mathit{publication}(z,x), \mathit{publication}(z,y).
	\end{align*}
	The hash table created by the algorithm contains the following mapping from constants to variables: \{ alice $\rightarrow x$, bob $\rightarrow y$, p1$\rightarrow z$, post\_quals $\rightarrow u$, assistant\_prof $\rightarrow v$\}.
	Note that there are two literals with relation {\it inPhase}, the first one created using mode definition \texttt{inPhase(+,-)} and the second one created using mode definition \texttt{inPhase(+,\#)}.
\end{example}

\begin{table}
	\centering
	\caption{Fragments of the UW-CSE database. }
	\begin{tabular} { l l }
		\hline
		student(alice) & professor(bob) \\
		student(john) & professor(mary) \\
		inPhase(alice,post\_quals) & hasPosition(bob,assistant\_prof) \\
		inPhase(john,post\_quals) & hasPosition(mary,associate\_prof) \\
		publication(p1,alice) & publication(p1,bob) \\
		publication(p2,john) & publication(p2,mary) \\
		\hline
	\end{tabular}
	\label{table:automode-example-database}
\end{table}

\subsection{Generalization}
\label{section:cator-generalization}

After building the bottom-clause associated with a given positive example, the algorithm generalizes the clause to cover more positive examples.
It uses the {\it asymmetric relative minimal generalization} ({\it armg}) operator to generalize clauses~\cite{progolem}. 
It performs a beam search to select the best 
clause generated after multiple applications of the {\it armg} operator.
More formally, given clause $C$, it randomly picks a subset $E^+_S$ of positive examples to generalize $C$.
For each example $e' \in E^+_S$, it uses the {\it armg} operator to generate a candidate clause $C'$, which is more general than $C$ and covers $e'$.
It then selects the highest scoring candidate clauses to keep in the beam and iterates until the clauses cannot be improved.
The score of clause is usually computed as the difference between the number of positive and negative examples
it covers.

We now explain the {\it armg} operator in detail.
Let $C$ be the bottom-clause associated with example $e$, relative to $I$.
Let $e'$ be another example.
$L_i$ is a {\it blocking atom} iff $i$ is the least value such that for all substitutions $\theta$ where 
$e' = T\theta$, 
the clause $C \theta = (T \leftarrow L_1, \cdots, L_i)\theta$ does not cover $e'$, relative to $I$~\cite{progolem}.
Given the bottom-clause $C$ and a positive example $e'$, {\it armg} drops all blocking atoms from the body of $C$
until $e'$ is covered. 
After removing a blocking atom, some literals in the body may not have any variable in common with the other literals in the body and head of the clause, i.e., they are not {\it head-connected}. 
{\it Armg} also drops those literals.
Because {\it armg} drops literals from the clause, it is guaranteed that the size of the clause reduces when doing generalization.
The bottom-clause is the most specific hypothesis that belongs to the hypothesis space.
Therefore, any generalization of it is also in the hypothesis space.

%% file: 4.System.tex
\section{Setting Language Bias Automatically}
\label{section:automode-system}

AutoBias leverages the information in the schema and content of the database to automatically generate predicate and mode definitions.
AutoBias reads and extracts the information about the schema 
of the underlying database from the relational database management system (RDBMS).
It then generates predicate and mode definitions in a preprocessing step.
AutoBias uses these definitions to learn the definition of some target relation.
The same predicate and mode definitions can be used to learn different target relations.

\input{4.System-PredicateDefinitions}

\input{4.System-ModeDefinitions}

%% file: 4.System-PredicateDefinitions.tex
\subsection{Generating Predicate Definitions}
\label{section:automode-predicate-definitions}

Let $R$ and $S$ be two relation symbols in the schema of the underlying database.
Let $R(e_1,\cdots,e_n)$ and $S(o_1,\cdots,o_m)$ be two atoms in a clause $C$.
Let $e_i$ be the term in attribute $R[A]$ and $o_j$ be the term in attribute $S[B]$, and let $e_i$ and $o_j$ be assigned the same variable or constant. That is, clause $C$ joins $R$ and $S$ on $A$ and $B$.
Clause $C$ is satisfiable only if these attributes share some values in the input database.
Typically, the more frequently used joins are the ones over the attributes that participate in inclusion dependencies (INDs), such as  foreign-key to primary-key referential constraints. 
AutoBias uses INDs in the input database to find which attributes, among all relations, share the same type.
Let $X$ and $Y$ be sets of attribute names in $R$ and $S$, respectively. Let $I_R$ and $I_S$ be the relations of $R$ and $S$ in the database. Relations $I_R$ and $I_S$ satisfy {\it exact IND} ({\it IND} for short) $R[X] \subseteq S[Y]$ if $\pi_{X}(I_R) \subseteq$ $\pi_{Y}(I_S)$.
If $X$ and $Y$ each contain only a single attribute, the IND is a {\it unary IND}.
Given IND $R[X] \subseteq S[Y]$ in a database, the database satisfies unary IND $R[A] \subseteq S[B]$, where $A \in X$ and $B \in Y$. 
INDs are normally stored in the schema of the database.
If they are not available in the schema, one can extract them from the database content. 
AutoBias uses the Binder algorithm~\cite{Papenbrock:2015:DCI:2752939.2752946} 
to discover INDs from the database and generates all unary INDs implied by them.
Binder efficiently discovers INDs by using a divide-and-conquer approach. First, it generates all unary candidate INDs. Second, it partitions the input dataset into small buckets that fit in main memory. Third, it loads each bucket into memory and validates the candidate INDs against the current bucket. The algorithm returns all INDs that pass all checks.

We have observed that in some cases using exact INDs is not enough for generating helpful predicate definitions. 
Consider two attributes $A_1$ and $A_2$, which contain values for domains $D_1$ and $D_2$, respectively.
There may be another attribute $A_3$ that contains some values from $D_1$ and some values from $D_2$. 
It makes sense to join attributes $A_1$ (or $A_2$) with $A_3$, as $A_1$ and $A_3$ contain values for domain $D_1$. However, exact INDs may not hold between $A_1$ (or $A_2$) and $A_3$.
An example of this scenario can be seen in the UW-CSE database, whose schema fragments are shown in Table~\ref{table:automode-uwcse}.
Consider the task of learning a definition for the relation {\it advisedBy(stud, prof)}, which indicates that the student {\it stud} is advised by professor {\it prof}.
A relational learning algorithm may learn the following Datalog program for the {\it advisedBy} relation:
\begin{align*}
\mathit{ad}&\mathit{visedBy}(x,y) \leftarrow  \mathit{student}(x), \mathit{professor}(y), \\ & \mathit{publication}(z,x),  \mathit{publication}(z,y)
\end{align*}
which indicates that a student is advised by a professor if they have been co-authors of a publication.
This definition requires joining relations {\it publication}, {\it student}, and {\it professor} on attributes {\it publication[author]}, {\it student[stud]}, and {\it professor[prof]}. 
However, the UW-CSE database does not satisfy INDs {\it publication[author]} $\subseteq$ {\it student[stud]} or {\it publication[author]} $\subseteq$ {\it professor[prof]} because {\it publication[author]} contains both students and professors.

To account for the issue described above, AutoBias also uses {\it approximate INDs} to assign types to attributes.
In an {\it approximate unary IND} $(R[A] \subseteq S[B], \alpha)$, one has to remove at least $\alpha$ fraction of the distinct values in $R[A]$ so that the database satisfies $R[A] \subseteq S[B]$ \cite{Abedjan:2015:PRD:2811716.2811766}. 
Approximate INDs are not usually maintained in a schema and are instead discovered from the database content.
We have implemented a program to extract approximate INDs from the database.
We use a relatively high error rate, 50\%, for the approximate INDs to allow for a flexible hypothesis space.

\begin{figure}
	\centering
	\includegraphics[width=0.7\columnwidth]{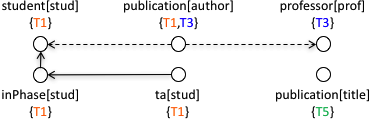}
	\caption{A fragment of the type graph for the UW-CSE dataset. Solid lines represent exact INDs and dashed lines represent approximate INDs.}
	\label{figure:graph}
\end{figure}

After discovering unary exact and approximate INDs,
AutoBias runs Algorithm~\ref{algorithm:typegraph} to generate a directed graph called {\it type graph}, which it then uses to assign types to attributes. 
First, it creates a graph whose nodes are attributes in the input schema and has an edge between each pair of attributes that participate in an exact or approximate IND.
Figure~\ref{figure:graph} shows an example of the type graph containing a subset of the attributes in the UW-CSE schema, where edges corresponding to exact and approximate INDs are shown by solid and dashed lines, respectively.
If there are both approximate INDs $(R[A] \subseteq S[B], \alpha_1)$ and $(S[B] \subseteq R[A], \alpha_2)$, AutoBias uses only the one with lower error rate.
The algorithm then assigns a new type to every node in the graph without any outgoing edges.
For example, it assigns new types \texttt{T1}, \texttt{T3}, and \texttt{T5} to {\it student[stud]}, {\it professor[prof]}, and {\it publication[title]}, respectively, in 
Figure~\ref{figure:graph}.
If there are cycles in the type graph, the algorithm assigns the same new type to all nodes in each cycle.
Next, it propagates the assigned type of each attribute to its neighbors in the reverse direction of edges in the graph until no changes are made to the graph. 
For example, in Figure~\ref{figure:graph}, the algorithm propagates type \texttt{T1} to {\it inPhase[stud]} and {\it ta[stud]} and attribute {\it publication[author]} inherits types \texttt{T1} and \texttt{T3} from {\it student[stud]} and {\it professor[prof]}, respectively.
Because the error rates of approximate INDs accumulate over multiple edges in the graph, AutoBias propagates types only once over edges that correspond to approximate INDs.

\begin{algorithm}
	\SetKwInOut{Input}{Input}
	\SetKwInOut{Output}{Output}
	\Input{Schema $\mS$ and all unary INDs $\Sigma$.}
	\Output{Type graph $G$.}
	
	create graph $G=(V,E)$ where $V$ contains a node for each attribute in the schema and $E=\varnothing$\;
	
	\ForEach{IND $R[A] \subseteq S[B] \in \Sigma$}
	{
		add edge $v \rightarrow u$ to $E$, where $v$ and $u$ correspond to attributes $R[A]$ and $S[B]$, respectively\;
	}
	\ForEach{node $u \in V$ without outgoing edges}
	{
		generate new type $T$ and set $types(u) = \{ T \}$\;
	}
	\ForEach{cycle $K \subseteq V$}
	{
		generate new type $T$ and set $types(u) = \{ T \}$ $\forall u \in K$\;
	}
	\Repeat{no changes in $G$}
	{
		\ForEach{$v \rightarrow u \in E$ where $types(u) \neq \varnothing$}
		{
			set $types(v) = types(v) \cup types(u)$\;
		}
	}
	return $G$\;	
	\caption{Algorithm to generate the type graph.}
	\label{algorithm:typegraph}
\end{algorithm}

Given the resulting graph, for each relation, AutoBias computes the Cartesian product of the types associated with its attributes. 
For each tuple in this Cartesian product, it produces a predicate definition for the relation.
For instance, given the type assignment in Figure~\ref{figure:graph}, AutoBias generates predicate definitions \texttt{publication(T5,T1)} and \texttt{publication(T5,T3)} for the {\it publication} relation.

%% file: 4.System-ModeDefinitions.tex
\subsection{Generating Mode Definitions}
\label{section:automode-variables-vs-constants}

AutoBias allows every attribute of each relation be a variable.
However, it forces at least one variable in an atom to be an existing variable, i.e., appears in previously added atoms, to avoid generating Cartesian products in the clause.
For each attribute $A$ in relation $R$, AutoBias generates a mode definition for $R$ where attribute $A$ is assigned the $+$ symbol and all other attributes are assigned the $-$ symbol.
Hence, all attributes are allowed to have new variables except the attribute with symbol $+$.
For instance, AutoBias generates the mode definitions \texttt{publication(+,-)} and \texttt{publication(-,+)} for relation {\it publication} in Table~\ref{table:automode-uwcse}.

AutoBias uses a hyper-parameter called {\it constant-threshold} to determine whether an attribute can be a constant.
The value for constant-threshold can take an absolute or a relative threshold.
If it is an absolute threshold, AutoBias allows an attribute to be a constant if the number of distinct values in the attribute is below the value of constant-threshold.
If it is a relative threshold, AutoBias allows an attribute to be a constant if the ratio of distinct values of the attribute to the total number of tuples in the relation is below the value of constant-threshold. 
This hyper-parameter must be tuned by the user. As it has a relatively intuitive meaning, it is easy to determine which values or ranges to experiment with.

For each relation $R$ in the database, AutoBias finds all attributes in $R$ that can be constants using the aforementioned rule.
Then, it computes the power set ${\bf M}$ of these attributes.
For each non-empty set $M \in {\bf M}$, AutoBias generates a new set of mode definitions where it assigns $+$ and $-$ symbols as described above, except for the attributes in $M$, which are assigned the $\#$ symbol.
For example, AutoBias finds that the number of values in attribute {\it phase} of relation {\it inPhase} in Table~\ref{table:automode-uwcse} is smaller than the input threshold.
Then, this attribute can be constant and AutoBias generates the mode definition \texttt{inPhase(+,\#)} for relation {\it inPhase}.

%% file: 5.Sampling.tex
\section{Improving Efficiency Through Sampling}
\label{section:automode-sampling}
Relational learning over large databases is generally time-consuming as the learner has to explore
numerous possible hypotheses whose tests of coverage take long.
For each bottom-clause, the algorithm has to include literals per tuples that are connected to the
positive example and also literals per tuples in the database that are connected to the current ones
in the bottom-clause. As numerous tuples across multiple tables may be connected via some join paths to a given positive example over a large database, it may create extremely long bottom-clauses. 
For example, a bottom-clause may contain tens of thousands of literals over a database with about a million tuples after a couple of iterations.
Since the created bottom-clauses contain many literals, it will be time-consuming to generalize the clauses by applying the {\it armg} operator multiple times.
Moreover, the algorithm has to check the number of positive and negative examples
covered by the generalized clause in each generalization step. 
As the clause has many literals, it will be time-consuming to test whether it covers an example.

Experts usually avoid this problem by familiarizing themselves with the underlying domain and database 
and creating a sufficiently restrictive language bias to limit the
hypothesis space and guide the learner to consider only the ones they deem promising.
Since AutoBias does not use the experts' intervention and guidance, it may produce language bias that 
does not sufficiently limit the hypothesis space. Thus, it may take several hours for the algorithm
to learn over a large database. 
In this section, we study sampling techniques that allow relational learning algorithms to learn efficiently
over large databases when their language bias may not be sufficiently restricted.
We first investigate the use of sampling in creating bottom-clauses and then study the 
application of sampling techniques in approximate coverage testing.

\subsection{Bottom-Clause Sampling}
A bottom-clause $C_e$ associated with an example $e$ is the most specific clause in the hypothesis space that covers $e$. The bottom-clause construction algorithm consists of two phases.
First, it finds all the information in $I$ relevant to $e$, denoted by $I_e$.
Then, given the information relevant to $e$, it creates the bottom-clause $C_e$ by converting tuples 
in $I_e$ to literals in the bottom-clause and replacing constants with fresh and proper variables.
The tuple set $I_e$ may be large if many tuples in $I$ are relevant to $e$, which in turn makes
$C_e$ too large.
To overcome this problem, one may use some sampling technique to obtain a smaller tuple set $I_e^s \subseteq I_e$
such that $I_e^s$ contains a subset of the  information in $I_e$.
Ideally, the subset $I_e^s$ contains predictive patterns that will allow the learning algorithm to learn an accurate definition.
Then, instead constructing $C_e$ using $I_e$, one may create a bottom-clause $C_e^s$  with fewer literals than those of $C_e$ from tuples in $I_e^s$. 
Clauses $C_e$ and $C_e^s$ have the same head-literal with the
information of the underlying example $e$ but the body of $C_e^s$ has fewer literals than that of $C_e$.

\subsubsection{Na\"ive Sampling}
\label{section:automode-naive-sampling}
Let $C_e$ be a bottom-clause associated with example $e$.
A {\it na\"ive sample} $C_e^s$ of clause $C_e$ is the clause obtained the following way.
Let $I_R$ be the set of tuples in relation $R$ that can be added to $I_e^s$.
The na\"ive sampling algorithm obtains a uniform and random sample $I_R^s$ of $I_R$ 
and adds only the tuples in $I_R^s$ to $I_e^s$.
Let the {\it inclusion probability} $p(t)$ of tuple $t \in I_e$ be the probability that $t$ is included in $I_e^s$.
In a uniform sample, every tuple in $I_R$ is sampled independently with the same inclusion probability, i.e., $\forall t \in I_R, p(t) = \frac{1}{|I_R|}$.

Existing relational learning algorithms, such as Progol~\cite{progol} and ProGolem~\cite{progolem,castor:SIGMOD17}, use this technique to build bottom-clauses. In this method, however, $I_e^s$ may {\it not} contain the predictive patterns in $I_e^s$, i.e., the set of tuples connected to $e$ in $I$.
For example, let the bottom-clause construction algorithm pick a set of tuples $J \subset I_e^s$ 
in an iteration. Let $t$ and $s$ be two tuples in a relation in $I$ that are connected via some joins to 
a single and a hundred tuples in $J$, respectively. Intuitively, the relationships between 
tuples in $J$ and $s$ is stronger than the ones between tuples in $J$ and $t$.
Thus, it is reasonable to include $s$ with a higher probability in $I_e^s$ than that of $t$.
Nevertheless, the na\"ive sampling method includes $t$ and $s$ with equal probabilities to $I_e^s$.
In the worst case, $s$ may not be connected to any tuples in the database in addition to the ones
that have been already included in the bottom-clause. In this case, the algorithm may return 
a bottom-clause with just a few literals, which in turn may cause the learning algorithm to output
a clause that is too general and not sufficiently informative.
Moreover, this method is biased toward tuples in relations with fewer tuples as it assigns higher 
inclusion probabilities to them.

\subsubsection{Random Sampling}
\label{section:automode-random-sampling}
To address the aforementioned shortcomings of the na\"ive sampling method, 
one may obtain a random sample of the literals in the body of $C_e$ to construct the body of $C_e^s$.
This method, however, faces two challenges.
As explained in Section~\ref{section:cator-bc}, each literal in $C_e$ is head-connected, which means that it is either connected to the head-literal of $C_e$ via some shared variables 
or it has some variables in common with other literals in the body of $C_e$ that are head-connected.
As explained in Section~\ref{section:cator-generalization}, a literal that does not meet these conditions, i.e., is 
not head-connected, will be automatically removed during generalization.
Moreover, that literal will not offer any useful information in the database related to the underlying positive example. 
If one selects literals from the body of $C_e$ uniformly at random, 
none of the selected literals may not be head-connected.
Thus, the learning algorithm may simply return an empty clause after the first step of generalization. 
Also, if most of the selected literals are not head-connected, 
the algorithm will eliminate most of the literals in $C_e^s$ after the first step of generalization.
Hence, the subsequent generalizations may not have sufficient or interesting information about the underlying example in the database to generalize and learn. 
One may not get a useful clause that contains predictive information in 
$C_e$ by simply uniformly and randomly sampling each literal in its body.
Thus, every literal in $C_e^s$ must also be head-connected.
Moreover, to create $C_e^s$, one has to obtain a random sample of $I_e$ to construct $I_e^s$.
To create a random sample of $I_e$, one may construct $I_e$ and then randomly sample sufficiently many of 
its tuples to construct $I_e^s$. 
But, as we explained earlier in this section, $I_e$ may be very time-consuming to construct and materialize 
for large databases. 

To address the aforementioned challenges, we should define a reasonable inclusion probability for each literal 
in $C_e$ and equivalently each tuple in $I_e$ for the random sample such that the sampled clause
does not contain literals that are not head-connected.
Furthermore, we should be able to compute these probabilities without computing and materializing $C_e$ and $I_e$.
Next, we precisely compute this inclusion probability without materializing $I_e$.
The {\it right semi-join} of relations $R_1$ and $R_2$ on attributes $A$ and $B$,
denoted as $R_1 \rtimes_{R_1.A = R_2.B} R_2$, is the set of tuples
in $R_2$ such that the values of their attribute $B$ are equal to the value of $A$ of 
at least one tuple in $R_1$ \cite{DBBook}. 
\begin{example}
\label{example:semijoin}
Consider relations $U_1(A,B)$ and $U_2=(A,C)$ such that
$U_1=$ $\{(a_1,b_1),(a_2,b_2), \ldots, (a_2, b_k)\}$ and $U_2=$ $\{(a_0,c_1),$ $(a_2,c_2), (a_1,c_3), \cdots,(a_1,c_m) \}$, we have $U_1 \rtimes_{U_1.A = U_2.A} U_2$ $=$ $\{(a_2,c_2),$ $(a_1,c_3), \cdots,(a_1,c_m) \}$.
\end{example}
\noindent
For brevity, we call right semi-join simply as semi-join
and show $R_1 \rtimes_{R_1.A = R_2.B} R_2$ as $R_1 \rtimes_{A,B} R_2$ unless otherwise noted.
The bottom-clause construction algorithm in Section~\ref{section:cator-bc} is in fact iteratively 
applying semi-joins to the database relations to add tuples to $I_e$ that are directly or indirectly 
connected to the positive example $e$ according to the mode and predicate definitions.
More precisely, for each pair of attributes of the same type $A$ and $B$ 
between the target relation $T$ and the relation $R$ in the background knowledge 
according to the predicate definitions, the bottom-clause construction algorithm will 
add the tuples of $\{e\} \rtimes R$ to $I_e$.
It then adds the tuples from another relation $S$ to $I_e$ using the semi-join of
$\{e\} \rtimes R$ $\rtimes S$. Generally, the algorithm computes 
$\Cup ( \rtimes_{1 \leq i \leq d-1} R_1 \rtimes \ldots \rtimes R_{1+i})$ in its $d$th 
iteration where $R_1= T$ and $\{ R_2, \ldots, R_d \}$ is a multi-set of possibly non-distinct relations in 
the background knowledge such that $R_i$ and $R_{i+1}$ have attributes of same type
according to the mode definitions.
Thus, we should efficiently compute a random sample of every
$R_1 \rtimes \ldots \rtimes R_{1+i}$.

To compute a random sample of $R_1 \rtimes_{A,B} R_2$, one should 
materialize $R_1 \rtimes_{A,B} R_2$ and then take a random sample of it. 
Nonetheless, this defeats the purpose of not computing $I_e$.
Another approach is to take independent random samples of $R_1$ and $R_2$ and semi-join them. 
However, the results may be empty and have very few tuples. 
Thus, it may take a long time to get a sufficiently large sample. 
Consider the relations $U_1$ and $U_2$ in Example~\ref{example:semijoin}.
It is very unlikely for a random sample of $U_1$ to contain a tuple whose value for $A$ is $a_1$
for a sufficiently large $k$. Also, a random sample of $U_2$ is unlikely to have a tuple whose
$A$ value is $a_2$ for large values of $m$. 
Thus, the semi-join of the random samples of $U_1$ and $U_2$ may be empty
for a reasonably large number of sampling rounds. 

Hence, we extend existing techniques for performing efficient 
sampling over joins~\cite{Chaudhuri1999OnRS,olkenThesis,Zhao2018RandomSO} to sample over semi-join $R_1 \rtimes_{A,B} R_2$ efficiently. 
Let $S_1$ be such a random sample of $R_1$.
The distributions of attribute values in tuples of $S_1$ are influenced by the ones of the tuples in $R_1$. 
For instance, a random sample of $U_1$ in Example~\ref{example:semijoin} contains mostly tuples whose 
values of attribute $A$ is $a_2$. But, the values of attribute $A$ of tuples in $U_1 \rtimes_{A,A} U_2$ 
are mostly $a_1$. Thus, one should accept the results of $S_1 \rtimes_{A,B} R_2$ based on the distribution of 
attribute values in $R_2$ to create a random sample of $R_1 \rtimes_{A,B} R_2$.
Furthermore, let the tuple $t \in R_2$ be the only tuple in $R_2$ whose value of attribute $B$ is $b$.
Let $b$ appear in the attribute $A$ of only a single tuple of $R_1$.
In this case, $t$ will be the only tuple in $R_1 \rtimes_{A,B} R_2$ whose value of $B$ is $b$.
Now, assume that $b$ appears in the attribute $A$ of more than a single tuple of $R_1$.
This will not change the number of tuples in $R_1 \rtimes_{A,B} R_2$ whose value for attribute $B$
is $b$. 
Thus, the distribution of values in $R_1 \rtimes_{A,B} R_2$ depends on the 
existence of values in $R_1[A]$ but does not change if their frequencies go beyond 1.
Therefore, one may randomly select only from values in the set of $\pi_{A} R_1$ and use it to  
compute a random sample of the semi-join.
Computing the distribution of values of $R_1 \rtimes_{A,B} R_2$ based on the existence of values in $R_1[A]$ instead of their frequencies is the only difference between random sampling over semi-joins as compared to random sampling over joins.

We adapt the extended Olken algorithm~\cite{olkenThesis} for performing random sampling over multi-way joins proposed by Zhao et al.~\cite{Zhao2018RandomSO} to work over semi-joins.
Our sampling algorithm over semi-join $R_1 \rtimes_{R_1.A = R_2.B} R_2$ is as follows. 
We first select a random value from all values of the set of $\pi_{A} R_1$ called $a$. 
Let $m_{R_2.B}(a)$ denote the frequency of $a$ in attribute $B$ of $R_2$.
Let $M_{R_2.B}$ is an upper bound on the frequency of each value of $B$ in $R_2$.
From all tuples in $R_2$ whose values of attribute $B$ is $a$, we select a tuple $t$ randomly.
We accept $t$ with the probability $p=$ $\frac{m_{R_2.B}(a)}{M_{R_2.B}}$ and reject it with 
$1-p$. We repeat this process from sampling a value from $\pi_{A} R_1$
from the beginning until a given number of tuples from $R_2$ are picked.
To compute the values of $m_{R_2.B}(a)$ and $M_{R_2.B}$ and find tuples of $R_2$ that match $a$
efficiently, we build indexes over the semi-join attributes \cite{olkenThesis,Chaudhuri1999OnRS,Zhao2018RandomSO}.
To compute the semi-join $R_1 \rtimes R_2 \ldots \rtimes R_n$, 
we compute the sample $S_2$ of $R_1 \rtimes R_2$ using the aforementioned algorithm.
Then, we compute the sample $S_3$ of $S_2 \rtimes R_3$ using this algorithm and 
continue the same process until the sample of semi-join $S_{n-1} \rtimes R_n$ is calculated.

\begin{proposition}
The aforementioned algorithm produces a random sample $R_1 \rtimes R_2 \ldots \rtimes R_n$.
\end{proposition}
\begin{proof}
The proof follows the one of random sampling over multi-way joins proposed by Olken~\cite{olkenThesis}.
\end{proof}
\noindent
The samples of some $S_i \rtimes R_{i+1}$, $1 < i < n $ might be empty as the values
in $S_i$ may not match any tuple in $R_{i+1}$. In this case, one has to repeat the sampling of
a preceding binary semi-join to get different values from the ones in $S_i$.
To avoid this problem, we take sufficiently larger number of samples than the desired final number of samples in each binary semi-join.

Given an input number of iterations $d$, the bottom-clause construction algorithm computes all semi-joins
of size up to $d$ and unions their output to construct $I_e$. To share computation between different samplings,
we organize all relations that will be semi-joined according to the predicate definitions 
in a {\it semi-join tree} $G$ of depth $d$.
Each node in $G$ is a relation symbol in the schema.
The root of $G$ represents the target relation symbol, $T$. 
Let $n_R$ be a node in $G$ that represents relation $R$.
A node $n_{R_1}$ in $G$ has a child $n_{R_2}$ if $R_1$ and $R_2$ can be semi-joined according to the mode
definitions. If the semi-join of $R_1$ and $R_2$ is $R_1 \rtimes_{A,B} R_2$, we place the label
$(A,B)$ on the edge from $n_{R_1}$ to $n_{R_2}$ in $G$.
Since a relation $R_2$ may appear at the right hand side of multiple semi-joins according to 
the mode definitions, $R_2$ may be represented by multiple distinct nodes in $G$.

Next, we apply the sampling algorithm following edges in $G$ starting from its root to generate the 
sample of $I_e$, $I_e^s$.
We consider the example $e$ as the only tuple of the relation of the root of $G$, which is sampled with 
probability of 1.
This enables us to share and reuse the random sample of a semi-join for the subsequent and longer ones.
After sampling the semi-join between a parent $n_{R_1}$ and one of its children $n_{R_2}$, 
we add the sampled tuples to $I_e^s$. We also use this set for the semi-join of $n_{R_2}$ and its children.
After constructing $I_e^s$, we create the bottom-clause $C_e^s$ according to $I_e^s$.
Different paths in $G$ may share some tuples. In this case, the union of randomly sampling from a set of 
relations is not exactly equivalent to random sampling over the union of the relations \cite{olkenThesis}.
We, however, make the simplifying assumption that they are equivalent to ensure sampling is efficient over
large databases. Otherwise, sampling will require considering the intersection of various semi-joins in $G$, 
which needs significantly more computations.

\subsubsection{Stratified Sampling}
\label{section:automode-stratified-sampling}
Random sampling may be biased toward relations and tuples that are strongly connected to other
relations and tuples in the database. Thus, it may miss some patterns in the data that effectively define
the training examples but are not sufficiently well-connected in the database \cite{DBLP:conf/icml/JensenN02}.
To address this problem, we propose a method that samples a diverse sample of 
tuples and relationships in the database to construct a sufficiently diverse sample $I_e^s$ of $I_es$ according to the mode and predicate definitions.
As explained in Section~\ref{section:automode-languagebias}, language bias sets two types of restrictions
on the patterns extracted from the data. It determines whether an attribute may be a variable or constants
in each literal of the bottom-clause and what join paths connect its literals.
We provide a sample that contains each possible variation of every literal  and 
ensures that the sampled bottom-clause covers all join paths that connect them according to the language bias.

Let $G$ be a semi-join tree defined in Section~\ref{section:automode-random-sampling} 
whose only tuple of its root node is example $e$.
Let $S$ be a relation that contains attribute $A$, where $A$ can appear as a constant according to the language bias and let $n_S$ is a node that represent $S$ in $G$.
We replace each $n_S$ with a set of new nodes each of which represent a relation that is a subset of $S$ 
with a distinct value for $S[A]$.
The parents of these nodes are the same as $n_S$.
Given a node $n_R$ in $G$, we define a {\it stratum} for each child of $n_R$.
Therefore, there is a stratum for each relation $S$ that can join with $R$ and, if $S$ contains an attribute $A$ that can be a constant, there is a stratum for each distinct value in $S[A]$.
A {\it stratified sample} $I_e^s$ of $I_e$ is a subset of $I_e$ that contains at least one tuple for each stratum in $G$.
A stratified sample $C_e^s$ of clause $C_e$ is the clause created from the stratified sample $I_e^s$ of $I_e$.

Algorithm~\ref{algorithm:bottom-clause-stratified-sample} depicts the bottom-clause construction algorithm using stratified sampling.
The algorithm traverses the semi-join tree $G$ in a depth-first manner. 
Once it reaches a given depth $d$, it computes the strata in the current relation, e.g., relation $S$.
If $S$ contains an attribute $A$ that can be constant according to the language bias, the algorithm creates a stratum for each distinct value for $S[A]$.
If $S$ does {\it not} contain attributes that can be constant according to the language bias, the only stratum is the set of all tuples in  $S$.
It then uniformly samples $s$ tuples for each stratum in $S$ and adds them to $I_e^s$.
Therefore, $I_e^s$ is the union of the all the  sampled strata in $S$.
When the algorithm backtracks to the parent relation $R$ of $S$, it adds all tuples in $R$ that join with the sampled tuples in $S$ to $I_e^s$.

\begin{algorithm}
	\SetKwProg{Fn}{Function}{:}{}
	\SetKwFunction{Strat}{Strat}
	\SetKwFunction{StratRec}{StratRec}
	
	\SetKwInOut{Input}{Input}
	\SetKwInOut{Output}{Output}
	\Input{example $e$, \# of iterations $d$, sample size $s$}
	\Output{bottom-clause $C_e$}
	
	$I_e^s = \{\}$\;
	
	\ForEach{attribute $A$ in $e$}{
		\ForEach{relation $R$ containing attribute $A$}{
			$I_e^s = I_e^s \cup \mathit{StratRec}(R, A, \{e[A]\}, 1, d, s)$
		}
	}
	
	$C_e^s=$ create clause from $e$ and $I_e^s$\;
	return $C_e^s$\;

	\SetKwProg{Pn}{Function}{:}{\KwRet}
	\Pn{\StratRec{$R$, $A$, $M$, $i$, $d$, $s$}}{
		$I_e^s = \{\}$\;
		$I_R = \sigma_{A \in M}(R)$\;
		\If{$i = d$ (last iteration)}{
			$I_e^s = I_e^s \cup \mathit{SampleStrata}(I_R,s)$\;
		} \Else{
			
			\ForEach{attribute $B$ in $R$}{
				\ForEach{relation $S$ containing attribute $B$}{
					$I_S = \mathit{StratRec}(S,B, \pi_B (I_R), i+1, d, s)$\;
					$I_e^s = I_e^s \cup (\sigma_{B \in \pi_B (I_S)} (I_R))$\;
				}
			}
		}
		
		return $I_e^s$\;
	}
	
	\caption{Bottom-clause construction algorithm using stratified sampling.}
	\label{algorithm:bottom-clause-stratified-sample}
\end{algorithm}

Since stratified sampling algorithm has to traverse and backtrack nodes in $G$ and perform corresponding operations, it may take longer than random sampling over a large database with a complex schema or 
language bias specifications. However, it does not need the precomputed statistics and indexes needed
to perform random sampling efficiently. It also may not face the problem of empty sampled relations
over long semi-joins. 

\subsection{Approximate Coverage Testing}
\label{section-approximate}
As explained in Section~\ref{section:cator-generalization}, during generalization, we have to compute the numbers
of positive and negative examples covered by a generalized clause to evaluate its quality.
One approach is to translate the clause to a Select-Project-Join SQL query and execute it over the underlying
database and examples. Nonetheless, these clauses may contain hundreds of literals in the several rounds of 
generalizations. Our empirical investigations show that it may take a long time to evaluate such SQL queries over a large database. 
Thus, we follow the approach used in relational learning algorithms and use 
$\theta$-subsumption to compute the coverage of candidate clauses \cite{progol,progolem,castor:SIGMOD17}.

In this approach, one builds a {\it ground bottom-clause} for each positive and negative example
using the bottom-clause construction algorithm in Section~\ref{section:cator-bc} in which constants are {\it not}
replaced with variables. 
A substitution $\theta$ replaces constants and variables in clause $C_1$ with a set of fresh constants or variables.
The resulting clause is denoted as $C_1 \theta$.
Clause $C$ $theta$-subsumes ground bottom-clause $G$ if and only if 
there is some substitution $theta$ such that $C \theta \subseteq G$, i.e., 
the set of literals in the body of $C \theta$ is a subset or equal to the set of literals in the body
of $G$.
To test whether a clause covers an example, we check if the clause 
subsumes the ground bottom-clause of the example. 
Since subsumption testing is NP-hard, we use approximation algorithm to test subsumption between long clauses \cite{progol}.
We use a subsumption engine to test coverage \cite{Kuzelka:2009:RSE:1497096.1497102}

Ideally, a ground bottom-clause $G_e$ for example $e$ must contain one literal per each tuple in the database that is connected to $e$ through some joins.
Otherwise, the $\theta$-subsumption test may declare that $C$ does {\it not} cover $e$ when $C$ actually covers $e$.
However, it may be time-consuming to check $\theta$-subsumption for clauses with many literals.
Since a learning algorithm performs numerous coverage testing during learning, it is essential to improve the time of coverage testing otherwise learning may take an extremely long time. 
Hence, we use the three aforementioned sampling techniques to generate ground bottom-clauses.
Given that the bottom-clause is built using sampling technique ${\bf S}$,  
we also use ${\bf S}$ to generate all ground bottom-clauses for learning.

%% file: 6.Experiments.tex
\section{Empirical Results}

\subsection{Datasets}
\label{section:setting}

\begin{table}
	\centering
	\caption{ Number of relations (\#R), tuples (\#T), positive examples (\#P), and negative examples (\#N) for each dataset. }
	\begin{tabular} { c c c c c }
		\hline
		Name & \#R & \#T & \#P & \#N \\
		\hline
		UW-CSE  & 9 & 1.8K & 102 & 204 \\ 
		HIV  & 5 & 7.9M & 2K & 4K \\ 
		IMDb  & 46 & 8.4M & 1.8K & 3.6K \\ 
		\hline
	\end{tabular}
	\label{table:automode-datasets}
\end{table}

We run experiments over three datasets whose information is shown in Table~\ref{table:automode-datasets}.

\begin{enumerate}[noitemsep, topsep=0pt, leftmargin=*]
\item {\bf UW-CSE}:
The UW-CSE database contains information about a computer science department.
We learn the target relation {\it advisedBy(stud, prof)}, which indicates that student {\it stud} is advised by professor {\it prof}.

\item {\bf HIV}:
The HIV database contains structural information about chemical compounds ({\it wiki.nci.nih.gov/display\\/NCIDTPdata}). 
We learn the target relation {\it antiHIV(comp)}, which indicates that compound with id {\it comp} has anti-HIV activity.
In this dataset, the positive examples are compounds known to have anti-HIV activity, while negative examples are compounds known to lack anti-HIV activity.

\item {\bf IMDb}:
The IMDb database ({\it imdb.com}) contains contains information about movies and people who make them.
We learn the target relation {\it dramaDirector(dir)}, which indicates that person with id {\it dir} has directed a drama movie.
\end{enumerate}

We compare the quality of the learned definitions using the metrics of {\it precision} and {\it recall} \cite{DeRaedt:2010:LRL:1952055}. 
Let the set of true positives for a Horn definition be the set of positive examples in the testing data that are covered by the Horn definition.  The precision of a Horn definition is the proportion of its true positives over all examples covered by the Horn definition. The recall of a Horn definition is the number of its true positives divided by the total number of positive examples in the testing data. 
Precision and recall are between 0 and 1, where an ideal definition delivers both precision and recall of 1.
F-measure is the weighted harmonic mean of the precision and recall.
We perform 10-fold cross validation for HIV and IMDb datasets and 5-fold cross validation for UW-CSE due to its relatively smaller size.
We evaluate precision, recall, and learning time, showing the average over the cross validation.

\subsection{Systems}

We implement AutoBias over {\it Castor}, an open source relational learning algorithm 
that is shown to be more effective that other available systems \cite{castor:SIGMOD17}. 
Castor is built on top of VoltDB, a main-memory relational database management system, {\it voltdb.com}.

We compare AutoBias against Castor and Aleph~\cite{aleph}. 
We use Castor with different ways of setting the language bias.
\begin{enumerate}[noitemsep, topsep=0pt,leftmargin=*]
	\item {\bf Castor-Baseline} assigns the same types to all attributes and allows every attribute to be a variable or a constant. 
	
	\item {\bf Castor-Baseline without constants} is the same as the baseline method, except that it does not allow any attribute to be a constant. 
	
	\item {\bf Castor-Manual tuning} uses the language bias written by an expert who has knowledge of the relational learning system and knows how to write predicate and mode definitions.
	The expert had to learn the schema and go through several trial and error phases by running the underlying learning system and observing its results to write the predicate and mode definitions.
	The expert wrote 19, 14, and 112 predicate and mode definitions for the UW-CSE, HIV, and IMDb databases, respectively. 
	
	\item {\bf Aleph} is a popular and public relational learning system, which as opposed to Castor does not use relational database systems to store and query the background knowledge and training data. 
	Similar to Auto-Bias, Aleph follows the sequential covering algorithm shown in Algorithm~\ref{algorithm:castor-covering}. However, Aleph follows a top-down approach.
	Aleph can emulate multiple relational learning algorithms.
	We configure Aleph to emulate FOIL \cite{Quinlan:FOIL,QuickFOIL}, which is a popular and well-known top-down relational learning algorithm.
	QuickFOIL is another implementation of FOIL that uses a relational database system to improve the running time 
	of FOIL \cite{QuickFOIL}. We, however, are not able to find a publicly available version of QuickFOIL.
	As any other relational learning algorithm, Aleph requires manual tuning to setup its language biases.
	We use the same predicate and mode definitions used for Castor-Manual tuning.

	\item {\bf AutoBias} generates predicate and mode definitions automatically, as described in Section~\ref{section:automode-system}. 
	The original databases do not contain INDs.
	Therefore, AutoBias calls the IND discovery tools explained in Section~\ref{section:automode-system}.
	The preprocessing step of AutoBias to extract INDs takes 1.2 seconds, 1.4 minutes, and 7.8 minutes over the UW-CSE, HIV, and IMDb databases, respectively.
	We set the constant-threshold hyper-parameter (Section~\ref{section:automode-variables-vs-constants}) to 5 for UW-CSE, 80 for HIV, and 400 for IMDb due to their different sizes.
\end{enumerate}

Over all settings of Castor and AutoBias, we build bottom-clauses and ground bottom-clauses using na\"ive sampling to make our results comparable to the ones of Castor.
Aleph also uses na\"ive sampling. We set the sampling rate to at most ten tuples per mode for each dataset.
In Section~\ref{section:automode-experiments-sampling}, we evaluate different sampling techniques.
We run experiments on a 2.3GHz Intel Xeon E5-2670 processor, running CentOS Linux 7.2 with 500GB of main memory.

\subsection{Manual Tuning Versus AutoBias}
\label{section:automode-experiments}
Table~\ref{table:automode-results} illustrates the results of our experiments. 

{\bf Castor-Baseline.}
Over the UW-CSE database, Castor is less accurate and efficient compared to other settings.
Over the HIV database, Castor obtains competitive precision and recall, but is significantly less efficient than manual tuning and AutoBias.
Over the IMDb database, Castor is killed by the kernel because of extreme use of resources.
By allowing every attribute to be a constant, every value in the database -- even if it has a non-predictive value -- may appear in a literal as a constant.
Therefore, the generated bottom-clause contains a large number of literals, many of which are not useful for learning a definition for the target relation. 
For instance, the first bottom-clause created when running over the IMDb databases contains on average 1255 literals.
Further, by assigning the same type to all attributes in all relations, it allows all relations to join with each other on any attribute, resulting in a long running time. 

{\bf Castor-Baseline without constants.}
Over the UW-CSE database, this setting is the most efficient, and obtains competitive precision and recall compared to manual tuning and AutoBias.
Over the HIV database, Castor does not terminate after 10 hours. In this case, because no constants are allowed, Castor is not able to find any definition that covers many positive examples. Therefore, Castor generates a bottom-clause for each positive example and tries to generalize it to cover more positive examples. This process is time-consuming.
Over the IMDb database, the perfect definition for the target relation {\it dramaDirector} contains a constant.
However, in this setting, constants are not allowed. Therefore, Castor learns other definitions which are significantly less accurate compared to manual tuning or AutoBias.

{\bf Castor-Manual tuning.}
Over the UW-CSE and IMDb databases, Castor with manual tuning results in the most effective definitions.
Over the HIV database, it obtains less effective results compared to the baseline or AutoBias.
Castor with manual is efficient over all datasets.
However, an expert had to spend significant amount of time tuning the language bias. Further, a non-expert user would not be able to specify this bias.

{\bf Aleph.} Since the top-down learning algorithm used by Aleph is generally biased toward learning
relatively short clauses, it is faster than other methods over UW-CSE and HIV datasets.
It takes Aleph longer than Castor with manual tuning and AutoBias to learn over IMDb, which is due to the fact that it does not use any 
underlying database system to access and query data. This approach does not scale for databases with 
numerous tuples, such as IMDb. Overall, Aleph with manual tuning delivers less effective 
definitions than those returned by Castor with manual tuning and AutoBias over all datasets.

{\bf AutoBias.}
In general, AutoBias is more effective than the baselines, and almost as effective as manual tuning.
AutoBias is less efficient than manual tuning.
Manually written predicate and mode definitions provide a more restricted hypothesis space than the ones generated by AutoBias, e.g., it allows less attributes to be constants or allows less attributes to be input variables.
Thus, Castor has to explore a larger hypothesis space when using AutoBias.
Nevertheless, the overhead in the running time is about 13 minutes for the HIV database and 4 minutes for the IMDb database, which is a reasonable overhead for saving an expert's time and the enterprise's financial resources that pay the machine learning expert.
There is no overhead over the UW-CSE database.
Hence, we argue that automating the generation of predicate and mode definitions with the cost of a modest overhead in performance is a reasonable trade-off.
Further, AutoBias enables non-experts to use relational learning systems more easily.

\begin{table*}
	\centering
	\caption{Results of learning relations over UW-CSE, IMDb, and HIV data (h=hours, m=minutes, s=seconds).}
	\begin{tabular} { c c c c c c c }
		\hline
		\multirow{2}{*}{Dataset} & \multirow{2}{*}{Measure} & \multirow{2}{*}{Castor-Baseline} & Castor-Baseline  & Castor-Manual & \multirow{2}{*}{Aleph}  & \multirow{2}{*}{AutoBias} \\
		& & & (w/o const.)  & tuning & &  \\

		\hline
		\multirow{4}{*}{UW-CSE} & Precision & 0.76 & 0.96  & 0.93 & 0.78  &  0.84 \\ 
		& Recall & 0.50 & 0.48 & 0.54 & 0.17 & 0.54 \\
		& F-measure & 0.60 & 0.64 & 0.68 & 0.27  & 0.64 \\
		& Time & 47s & 6.6s & 11s & 3.5s & 24.4s 
		\\
		\hline
		
		\multirow{4}{*}{IMDb} & Precision & - & 0.68 & 1 & 0.66 & 1  \\
		& Recall & - & 0.51  & 0.99 & 0.44 & 0.99 \\
		& F-measure & - & 0.58  & 0.99 & 0.52 & 0.99 \\
		& Time & - & 9.2h  & 2.7m & 6.4m & 3.21m \\	
		\hline
				
		\multirow{4}{*}{HIV} & Precision & 0.80 & -  & 0.74 & 0.72 & 0.80  \\
		& Recall & 0.83 & -  & 0.84 & 0.69 & 0.85 \\
		& F-measure & 0.81 & -  & 0.78 & 0.70 & 0.82 \\
		& Time & 59.7m & $>$10h  & 22.6m & 6.2m & 35.1m \\
%
	\end{tabular}
	\label{table:automode-results}
	\vspace{-10pt}
\end{table*}
\vspace{-10pt}
\begin{table*}
	\centering
	\caption{Results over UW-CSE, HIV, and IMDb data with different sampling techniques (m=minutes), s=seconds.}
	\vspace{-10pt}
	\begin{tabular} { c c c c c c c }
		\hline
		\multicolumn{5}{c}{}  \\
		Dataset & Measure &  AutoBias-Na\"ive & \multicolumn{2}{c}{AutoBias-Random}  & \multicolumn{2}{c}{AutoBias-Stratified}  \\
		 & & & Mean & Variance & Mean & Variance \\
		\hline
		
		\multirow{4}{*}{UW-CSE} & Precision & 0.84 & 0.81 & 0.06 & 0.78 & 0.05 \\
		& Recall & 0.54 & 0.52  & 0.04 &  0.44 & 0.04 \\
		& F-measure & 0.64 & 0.61 & 0.04 &  0.54 & 0.05 \\
		& Time & 24.4s & 50.23s  & 27.84s &  37.86s & 13.76s \\
		\hline		
		
		\multirow{4}{*}{IMDb} & Precision & 1 & 0.99 & 0.004 &  0.99 & 0.006\\
		& Recall & 0.99 & 0.99  & 0.001 &  0.99 & 0.001\\
		& F-measure & 0.99 & 0.99 & 0.002 &  0.99 & 0.002\\
		& Time & 3.21m & 3.13m  & 0.27m &  4.05m & 0.09m\\
		\hline
		
		\multirow{4}{*}{HIV} & Precision & 0.80 & 0.79 & 0.04 & 0.83 & 0.014 \\
		& Recall & 0.85 & 0.87 & 0.02 & 0.76 & 0.009 \\
		& F-measure & 0.82 & 0.83 & 0.02 & 0.79 & 0.005 \\
		& Time & 35.1m & 21.87m  & 5.12m & 34.16m & 0.889m \\
		\hline
		
	\end{tabular}
	\label{table:automode-sampling-results}
\end{table*}

\subsection{Evaluating Sampling Techniques}
\label{section:automode-experiments-sampling}
In this section, we empirically evaluate the sampling techniques presented in Section~\ref{section:automode-sampling}.
We have implemented these sampling techniques by modifying the modules in charge of generating the bottom-clause construction and coverage testing in Castor. We use three versions of AutoBias. Each version uses a different sampling technique for bottom-clause construction: {\bf AutoBias-Na\"ive} uses na\"ive sampling, as explained in Section~\ref{section:automode-naive-sampling}, {\bf AutoBias-Random} uses random sampling, as explained in Section~\ref{section:automode-random-sampling}. {\bf AutoBias-Stratified} uses stratified sampling, as explained in Section~\ref{section:automode-stratified-sampling}. Each sampling method is used for both bottom-clause construction
and coverage testing. We have also built the necessary indexes for random sampling over 
relations in VoltDB. 
We use the same sampling rate of at most 10 tuples per each mode for all sampling methods and all datasets.

Table~\ref{table:automode-sampling-results} shows the effectiveness and efficiency of learning over 
using the aforementioned sampling techniques over 
the UW-CSE, IMDb, and HIV dataset.
We have run random and stratified sampling methods over 
each dataset five times and computed the average and variance 
of the resulting runs.

\subsubsection{Effectiveness}
AutoBias-Na\"ive and AutoBias-Random both achieve the highest F-measure and running time over
UW-CSE dataset. Due to the small size of the data and schema of UW-CSE, 
AutoBias-Na\"ive is able to create a sufficiently representative sample of the data and 
learn an effective definition over this dataset. 
AutoBias-Random learns a definition as effective as AutoBias-Na\"ive, which
indicates that it is able to generate a representative sample over this small dataset.
Since the data and its schema is not very large and diverse, AutoBias-Stratified does not offer any 
advantage over AutoBias-Random and returns a less effective definition than that of AutoBias-Random.
All methods return the effective definition for the IMDb database. 
Our observations indicate that when there is a relatively short definition that exactly 
represent the training data all methods can find an effective definition for the target relation.

AutoBias-Random obtains a higher precision and recall than AutoBias-Na\"ive over the HIV dataset.
In fact, AutoBias-Na\"ive obtains the lowest precision and recall among all methods over this dataset.
AutoBias-Stratified achieves a higher F-measure than that of AutoBias-Random over this dataset.
In particular, it delivers a considerably higher recall than the one of AutoBias-Random for this dataset.
This dataset is large and has a relatively complex schema with significant diversities in 
values and relationships. Moreover, its target relation is complex and there is not any 
definition with a reasonably many literals and clauses that covers all positive examples and 
does not cover any negative ones.
AutoBias-Stratified is able to create a sample that captures the diverse patterns in the data more than the one constructed by AutoBias-Random. Thus, its learned clauses are more helpful in explaining the patterns
in the training examples.
For example, in the HIV database, compounds contain hundreds of atoms. 
Some atoms are common elements, e.g., Hydrogen, while other atoms are rare elements, e.g., Lithium.
AutoBias-Stratified is able to explore join paths that lead to all types of elements in a compound.
Therefore, the bottom-clauses generated by AutoBias-Stratified contain diverse information, which allows it to learn better definitions. This helps AutoBias-Stratified to deliver a higher recall.

Overall, AutoBias-Random is able to consistently learn more effective or as effective definitions than
those of AutoBias-Na\"ive over all datasets. Its improvement over AutoBias-Na\"ive is particularly 
more considerable over large datasets with complex target relations. 
AutoBias-Stratified performs less effective than other approaches for small datasets or
the ones where there is a relatively short definition, e.g., SQL query, to explain the training data
fully. It, however, learns a more effective definition than others when the data is large and the target
relation is complex. Thus, one may safely use AutoBias-Random over datasets where one is not sure
about the diversity of the data and complexity of the target relation instead of 
AutoBias-Na\"ive. One may use AutoBias-Stratified when the target relation is complex and learning is done over 
a large dataset to improve the effectiveness of the definitions provided by AutoBias-Random.

\subsubsection{Efficiency}

AutoBias-Na\"ive is faster than AutoBias-Random over UW-CSE due to the overheads of selecting a random tuples over
various semi-joins. AutoBias-Na\"ive has a minimal overhead as 
it simply samples tuples uniformly in each relation.
Similarly, AutoBias-Stratified is less efficient than AutoBias-Na\"ive over UW-CSE for the same reasons.

Interestingly, AutoBias-Random improves the running time of AutoBias-Na\"ive over IMDb and HIV.
The efficiency is affected not only by the time taken to build a bottom-clause, but also by the speed in which effective definitions are discovered.
AutoBias-Random uses the covering approach (Algorithm~\ref{algorithm:castor-covering}).
Each time that a learned clause satisfies the minimum criterion, 
it is added to the definition and all positive examples covered by the clause are removed. 
If clauses that cover many positive examples are discovered early, then less positive examples must be used to generate candidate definitions. Therefore, the learning process will take less time.
AutoBias-Random constructs more representative samples than those of AutoBias-Na\"ive, therefore,
it creates bottom-clauses and consequently generalized clauses that 
cover more positive examples than the ones created by AutoBias-Na\"ive.
AutoBias-Stratified also constructs bottom-clauses and generalized clauses that cover more positive examples
than the ones covered by the clauses of AutoBias-Na\"ive. However, it is less efficient than AutoBias-Random
over all datasets due to the overhead of its backtracking.

Hence, AutoBias-Random is able to improve the running time of learning over sufficiently large database.
Using this approach, the running of AutoBias is equal to or less than the ones of Castor with manual tuning.
AutoBias-Stratified may be more efficient than AutoBias-Na\"ive if the underlying dataset is sufficiently large 
and the target relation is complex enough.
Both methods have low variance over all datasets but UWCSE, which is due its small size.

%% file: 7.Related.tex
\section{Related Work}
Recently, there has been a growing interest in developing relational learning algorithms that scale to large databases in both the database and machine learning communities \cite{amie-plus,castor:SIGMOD17,join:crossmine,QuickFOIL,S1}. 
The structure learning methods, such as the ones used to learn the structure of MLNs, use relational learning methods to learn 
the structure of their rules and templates \cite{Richardson:2006:MLN}. These systems usually use algorithms such as the ones used in Aleph to learn the structure of
their rules\cite{Richardson:2006:MLN,Quinlan:FOIL}. 
They may use the accuracy of the learned rules delivered by the relational 
learning algorithm as a weight for their rules for their probabilistic logical model.
There are, however, other methods to assign weights to the learned rules in frameworks, such as MLNs. 
For example, as they aim at learning all reliable rules, their structure learning methods may compute the weight of a rule
using the weights of other rules to create a consistent model \cite{Getoor:SRLBook}. 

There has been interest in reducing the user input in relational learning systems. 
The work in~\cite{Mccreath95extractionof} is similar to ours, where the goal is to induce predicate and mode definitions from data. 
Their algorithm assigns the same type to two attributes if there is an overlap of at least one element. This may result in an under-restricted search space because it may assign the same type to multiple attributes.
To generate mode definitions, their algorithm uses functional dependencies (FD) in a relation that involves all attributes in the relation. Then, it generates a mode definition for each FD, where it forces the attributes in the left-hand side of the FD to be existing variables.
This may result in an over-restricted hypothesis space, as in some cases, it may be useful to  consider attributes in the right-hand side attributes as input variables.
Moreover, it does not leverage sampling techniques to improve the efficiency and effectiveness of 
learning over large databases.

It is well-established that one of the important challenges of using a 
learning system is to set its hyper-parameters \cite{Hayes:2017:UFA:3148011.3148027}.
In particular, there has been some attempts to make the specification of language bias easier for the experts.
Some researchers provide the domain expert with a graphical representation of 
the underlying schema of the data, akin to an entity-relationship diagram, so she can specify the mode
and predicate definitions more easily \cite{Hayes:2017:UFA:3148011.3148027}.
Another approach is to ask the expert to provide some examples and advise in form of logical theories  
 \cite{Walker:2010:AIS:2022735.2022765}.
The system then constructs mode definitions from these logical statements.
All of these systems require heavy experts' intervention. Our goal, however, is to eliminate experts'
involvement in specifying language bias by leveraging the database constraints and patterns in the data.

These algorithms must restrict the hypothesis space through declarative bias or use a relatively 
restricted data model.
For instance, QuickFOIL~\cite{QuickFOIL} provides an in-RDBMS implementation of a modified version of FOIL. 
FOIL takes as input syntactic bias expressing the same information as predicate and mode definitions.
AMIE+~\cite{amie-plus} learns rules efficiently from RDF-style knowledge bases.
As AutoBias, AMIE+ forces at least one variable in an atom to be an existing variable.
However, RDF-style knowledge bases contain only binary relations, a more restricted data model than 
the one used in general relational learning as far as mode definitions are concerned. 
Our system extends this line of research by making relational learning both scalable and usable for the end-user.

Sampling has been extensively studied in the context of non-relational learning, e.g., stochastic gradient decent, \cite{MachineLearning:Mitchell,1424458,Bhojanapalli:2015:TLA:2722129.2722191,Jaiswal:2014:SDB:2644691.2644757}.
Nevertheless, to the best of our knowledge, efficient and effective sampling techniques have been largely unexplored for relational learning systems beyond na\"ive sampling \cite{progol,progolem}. 
Jensen et al. have pointed out the biases associated with 
selecting features over relational data by over-reliance on strongly connected relations \cite{DBLP:conf/icml/JensenN02}.

%% file: 7.Conclusion.tex
\section{Conclusion}
We have proposed AutoBias, a system that automatically induces the language bias used by relational learning algorithms. Our empirical studies indicate that AutoBias delivers a comparable learning effectiveness to 
the systems where the language bias is specified by experts. 

%% file: draft.bbl

\begin{thebibliography}{41}


\ifx \showCODEN    \undefined \def \showCODEN     #1{\unskip}     \fi
\ifx \showDOI      \undefined \def \showDOI       #1{#1}\fi
\ifx \showISBNx    \undefined \def \showISBNx     #1{\unskip}     \fi
\ifx \showISBNxiii \undefined \def \showISBNxiii  #1{\unskip}     \fi
\ifx \showISSN     \undefined \def \showISSN      #1{\unskip}     \fi
\ifx \showLCCN     \undefined \def \showLCCN      #1{\unskip}     \fi
\ifx \shownote     \undefined \def \shownote      #1{#1}          \fi
\ifx \showarticletitle \undefined \def \showarticletitle #1{#1}   \fi
\ifx \showURL      \undefined \def \showURL       {\relax}        \fi
\providecommand\bibfield[2]{#2}
\providecommand\bibinfo[2]{#2}
\providecommand\natexlab[1]{#1}
\providecommand\showeprint[2][]{arXiv:#2}

\bibitem[\protect\citeauthoryear{Abedjan, Golab, and Naumann}{Abedjan
  et~al\mbox{.}}{2015}]%
        {Abedjan:2015:PRD:2811716.2811766}
\bibfield{author}{\bibinfo{person}{Ziawasch Abedjan}, \bibinfo{person}{Lukasz
  Golab}, {and} \bibinfo{person}{Felix Naumann}.}
  \bibinfo{year}{2015}\natexlab{}.
\newblock \showarticletitle{Profiling relational data: a survey}.
\newblock \bibinfo{journal}{{\em The VLDB Journal\/}}  \bibinfo{volume}{24}
  (\bibinfo{year}{2015}), \bibinfo{pages}{557--581}.
\newblock


\bibitem[\protect\citeauthoryear{Abiteboul, Hull, and Vianu}{Abiteboul
  et~al\mbox{.}}{1994}]%
        {AliceBook}
\bibfield{author}{\bibinfo{person}{Serge Abiteboul}, \bibinfo{person}{Richard
  Hull}, {and} \bibinfo{person}{Victor Vianu}.}
  \bibinfo{year}{1994}\natexlab{}.
\newblock \bibinfo{booktitle}{{\em {Foundations of Databases: The Logical
  Level}}}.
\newblock \bibinfo{publisher}{Addison-Wesley}.
\newblock


\bibitem[\protect\citeauthoryear{Abouzeid, Angluin, Papadimitriou, Hellerstein,
  and Silberschatz}{Abouzeid et~al\mbox{.}}{2013}]%
        {Abouzied:PODS:13}
\bibfield{author}{\bibinfo{person}{Azza Abouzeid}, \bibinfo{person}{Dana
  Angluin}, \bibinfo{person}{Christos~H. Papadimitriou},
  \bibinfo{person}{Joseph~M. Hellerstein}, {and} \bibinfo{person}{Abraham
  Silberschatz}.} \bibinfo{year}{2013}\natexlab{}.
\newblock \showarticletitle{Learning and verifying quantified boolean queries
  by example}. In \bibinfo{booktitle}{{\em PODS}}.
\newblock


\bibitem[\protect\citeauthoryear{Bhojanapalli, Jain, and Sanghavi}{Bhojanapalli
  et~al\mbox{.}}{2015}]%
        {Bhojanapalli:2015:TLA:2722129.2722191}
\bibfield{author}{\bibinfo{person}{Srinadh Bhojanapalli},
  \bibinfo{person}{Prateek Jain}, {and} \bibinfo{person}{Sujay Sanghavi}.}
  \bibinfo{year}{2015}\natexlab{}.
\newblock \showarticletitle{Tighter Low-rank Approximation via Sampling the
  Leveraged Element}. In \bibinfo{booktitle}{{\em Proceedings of the
  Twenty-sixth Annual ACM-SIAM Symposium on Discrete Algorithms}} {\em
  (\bibinfo{series}{SODA '15})}. \bibinfo{pages}{902--920}.
\newblock


\bibitem[\protect\citeauthoryear{Chaudhuri, Motwani, and Narasayya}{Chaudhuri
  et~al\mbox{.}}{1999}]%
        {Chaudhuri1999OnRS}
\bibfield{author}{\bibinfo{person}{Surajit Chaudhuri}, \bibinfo{person}{Rajeev
  Motwani}, {and} \bibinfo{person}{Vivek~R. Narasayya}.}
  \bibinfo{year}{1999}\natexlab{}.
\newblock \showarticletitle{On Random Sampling over Joins}. In
  \bibinfo{booktitle}{{\em SIGMOD Conference}}.
\newblock


\bibitem[\protect\citeauthoryear{De~Raedt}{De~Raedt}{2010}]%
        {DeRaedt:2010:LRL:1952055}
\bibfield{author}{\bibinfo{person}{Luc De~Raedt}.}
  \bibinfo{year}{2010}\natexlab{}.
\newblock \bibinfo{booktitle}{{\em Logical and Relational Learning\/}
  (\bibinfo{edition}{1st} ed.)}.
\newblock \bibinfo{publisher}{Springer Publishing Company, Incorporated}.
\newblock
\showISBNx{3642057489, 9783642057489}


\bibitem[\protect\citeauthoryear{Domingos}{Domingos}{2018}]%
        {10.1145/3183713.3199515}
\bibfield{author}{\bibinfo{person}{Pedro Domingos}.}
  \bibinfo{year}{2018}\natexlab{}.
\newblock \showarticletitle{Machine Learning for Data Management: Problems and
  Solutions}. In \bibinfo{booktitle}{{\em Proceedings of the 2018 International
  Conference on Management of Data}} {\em (\bibinfo{series}{SIGMOD ?18})}.
  \bibinfo{publisher}{Association for Computing Machinery},
  \bibinfo{address}{New York, NY, USA}, \bibinfo{pages}{629}.
\newblock
\showISBNx{9781450347037}
\showDOI{%
\url{https://doi.org/10.1145/3183713.3199515}}


\bibitem[\protect\citeauthoryear{Dundar, Krishnapuram, Bi, and Rao}{Dundar
  et~al\mbox{.}}{2007}]%
        {10.5555/1625275.1625397}
\bibfield{author}{\bibinfo{person}{Murat Dundar}, \bibinfo{person}{Balaji
  Krishnapuram}, \bibinfo{person}{Jinbo Bi}, {and} \bibinfo{person}{R.~Bharat
  Rao}.} \bibinfo{year}{2007}\natexlab{}.
\newblock \showarticletitle{Learning Classifiers When the Training Data is Not
  IID}. In \bibinfo{booktitle}{{\em Proceedings of the 20th International Joint
  Conference on Artifical Intelligence}} {\em (\bibinfo{series}{IJCAI?07})}.
  \bibinfo{publisher}{Morgan Kaufmann Publishers Inc.}, \bibinfo{address}{San
  Francisco, CA, USA}, \bibinfo{pages}{756?761}.
\newblock


\bibitem[\protect\citeauthoryear{Evans and Grefenstette}{Evans and
  Grefenstette}{2018}]%
        {Evans2018LearningER}
\bibfield{author}{\bibinfo{person}{Richard Evans} {and} \bibinfo{person}{Edward
  Grefenstette}.} \bibinfo{year}{2018}\natexlab{}.
\newblock \showarticletitle{Learning Explanatory Rules from Noisy Data}.
\newblock \bibinfo{journal}{{\em J. Artif. Intell. Res.\/}}
  \bibinfo{volume}{61} (\bibinfo{year}{2018}), \bibinfo{pages}{1--64}.
\newblock


\bibitem[\protect\citeauthoryear{Gal{\'a}rraga, Teflioudi, Hose, and
  Suchanek}{Gal{\'a}rraga et~al\mbox{.}}{2015}]%
        {amie-plus}
\bibfield{author}{\bibinfo{person}{Luis Gal{\'a}rraga},
  \bibinfo{person}{Christina Teflioudi}, \bibinfo{person}{Katja Hose}, {and}
  \bibinfo{person}{Fabian~M. Suchanek}.} \bibinfo{year}{2015}\natexlab{}.
\newblock \showarticletitle{Fast rule mining in ontological knowledge bases
  with {AMIE+}}.
\newblock \bibinfo{journal}{{\em The VLDB Journal\/}}  \bibinfo{volume}{24}
  (\bibinfo{year}{2015}), \bibinfo{pages}{707--730}.
\newblock


\bibitem[\protect\citeauthoryear{GarciaMolina, Ullman, and Widom}{GarciaMolina
  et~al\mbox{.}}{2008}]%
        {DBBook}
\bibfield{author}{\bibinfo{person}{Hector GarciaMolina}, \bibinfo{person}{Jeff
  Ullman}, {and} \bibinfo{person}{Jennifer Widom}.}
  \bibinfo{year}{2008}\natexlab{}.
\newblock \bibinfo{booktitle}{{\em {Database Systems: The Complete Book}}}.
\newblock \bibinfo{publisher}{{Prentice Hall}}.
\newblock


\bibitem[\protect\citeauthoryear{Getoor and Taskar}{Getoor and Taskar}{2007}]%
        {Getoor:SRLBook}
\bibfield{author}{\bibinfo{person}{Lise Getoor} {and} \bibinfo{person}{Ben
  Taskar}.} \bibinfo{year}{2007}\natexlab{}.
\newblock \bibinfo{booktitle}{{\em {Introduction to Statistical Relational
  Learning}}}.
\newblock \bibinfo{publisher}{{MIT Press}}.
\newblock


\bibitem[\protect\citeauthoryear{Hayes, Das, Odom, and Natarajan}{Hayes
  et~al\mbox{.}}{2017}]%
        {Hayes:2017:UFA:3148011.3148027}
\bibfield{author}{\bibinfo{person}{Alexander~L. Hayes}, \bibinfo{person}{Mayukh
  Das}, \bibinfo{person}{Phillip Odom}, {and} \bibinfo{person}{Sriraam
  Natarajan}.} \bibinfo{year}{2017}\natexlab{}.
\newblock \showarticletitle{User Friendly Automatic Construction of Background
  Knowledge: Mode Construction from ER Diagrams}. In \bibinfo{booktitle}{{\em
  Proceedings of the Knowledge Capture Conference}} {\em
  (\bibinfo{series}{K-CAP 2017})}. \bibinfo{publisher}{ACM},
  \bibinfo{address}{New York, NY, USA}, Article \bibinfo{articleno}{30},
  \bibinfo{numpages}{8}~pages.
\newblock
\showISBNx{978-1-4503-5553-7}
\showDOI{%
\url{https://doi.org/10.1145/3148011.3148027}}


\bibitem[\protect\citeauthoryear{Jaiswal, Kumar, and Sen}{Jaiswal
  et~al\mbox{.}}{2014}]%
        {Jaiswal:2014:SDB:2644691.2644757}
\bibfield{author}{\bibinfo{person}{Ragesh Jaiswal}, \bibinfo{person}{Amit
  Kumar}, {and} \bibinfo{person}{Sandeep Sen}.}
  \bibinfo{year}{2014}\natexlab{}.
\newblock \showarticletitle{A Simple D2-Sampling Based PTAS for k-Means and
  Other Clustering Problems}.
\newblock \bibinfo{journal}{{\em Algorithmica\/}} \bibinfo{volume}{70},
  \bibinfo{number}{1} (\bibinfo{date}{Sept.} \bibinfo{year}{2014}),
  \bibinfo{pages}{22--46}.
\newblock
\showISSN{0178-4617}


\bibitem[\protect\citeauthoryear{Jensen and Neville}{Jensen and
  Neville}{2002}]%
        {DBLP:conf/icml/JensenN02}
\bibfield{author}{\bibinfo{person}{David~D. Jensen} {and}
  \bibinfo{person}{Jennifer Neville}.} \bibinfo{year}{2002}\natexlab{}.
\newblock \showarticletitle{Linkage and Autocorrelation Cause Feature Selection
  Bias in Relational Learning}. In \bibinfo{booktitle}{{\em Machine Learning,
  Proceedings of the Nineteenth International Conference {(ICML} 2002),
  University of New South Wales, Sydney, Australia, July 8-12, 2002}}.
  \bibinfo{pages}{259--266}.
\newblock


\bibitem[\protect\citeauthoryear{Kalashnikov, Lakshmanan, and
  Srivastava}{Kalashnikov et~al\mbox{.}}{2018}]%
        {Kalashnikov:2018:FFQ:3183713.3183727}
\bibfield{author}{\bibinfo{person}{Dmitri~V. Kalashnikov},
  \bibinfo{person}{Laks~V.S. Lakshmanan}, {and} \bibinfo{person}{Divesh
  Srivastava}.} \bibinfo{year}{2018}\natexlab{}.
\newblock \showarticletitle{FastQRE: Fast Query Reverse Engineering}. In
  \bibinfo{booktitle}{{\em Proceedings of the 2018 International Conference on
  Management of Data}} {\em (\bibinfo{series}{SIGMOD '18})}.
  \bibinfo{publisher}{ACM}, \bibinfo{address}{New York, NY, USA},
  \bibinfo{pages}{337--350}.
\newblock
\showISBNx{978-1-4503-4703-7}
\showDOI{%
\url{https://doi.org/10.1145/3183713.3183727}}


\bibitem[\protect\citeauthoryear{Kimmig, Poole, and Pujara}{Kimmig
  et~al\mbox{.}}{[n. d.]}]%
        {StarAIW}
\bibfield{author}{\bibinfo{person}{Angelika Kimmig}, \bibinfo{person}{David
  Poole}, {and} \bibinfo{person}{Jay Pujara}.} \bibinfo{year}{[n.
  d.]}\natexlab{}.
\newblock \showarticletitle{Statistical Relational AI (StarAI) WorkShop}.
\newblock


\bibitem[\protect\citeauthoryear{Kraska et~al\mbox{.}}{Kraska
  et~al\mbox{.}}{2013}]%
        {MLBase:CIDR}
\bibfield{author}{\bibinfo{person}{Tim Kraska} {et~al\mbox{.}}}
  \bibinfo{year}{2013}\natexlab{}.
\newblock \showarticletitle{{MLbase: A Distributed Machine-learning System}}.
  In \bibinfo{booktitle}{{\em {CIDR}}}.
\newblock


\bibitem[\protect\citeauthoryear{{Krishnapuram}, {Carin}, {Figueiredo}, and
  {Hartemink}}{{Krishnapuram} et~al\mbox{.}}{2005}]%
        {1424458}
\bibfield{author}{\bibinfo{person}{B. {Krishnapuram}}, \bibinfo{person}{L.
  {Carin}}, \bibinfo{person}{M.~A.~T. {Figueiredo}}, {and}
  \bibinfo{person}{A.~J. {Hartemink}}.} \bibinfo{year}{2005}\natexlab{}.
\newblock \showarticletitle{Sparse multinomial logistic regression: fast
  algorithms and generalization bounds}.
\newblock \bibinfo{journal}{{\em IEEE Transactions on Pattern Analysis and
  Machine Intelligence\/}} \bibinfo{volume}{27}, \bibinfo{number}{6}
  (\bibinfo{year}{2005}), \bibinfo{pages}{957--968}.
\newblock


\bibitem[\protect\citeauthoryear{Kumar, Naughton, and Patel}{Kumar
  et~al\mbox{.}}{2015}]%
        {Kumar:2015:LGL:2723372.2723713}
\bibfield{author}{\bibinfo{person}{Arun Kumar}, \bibinfo{person}{Jeffrey
  Naughton}, {and} \bibinfo{person}{Jignesh~M. Patel}.}
  \bibinfo{year}{2015}\natexlab{}.
\newblock \showarticletitle{Learning Generalized Linear Models Over Normalized
  Data}. In \bibinfo{booktitle}{{\em SIGMOD}}.
\newblock


\bibitem[\protect\citeauthoryear{Kuzelka and Zelezn{\'y}}{Kuzelka and
  Zelezn{\'y}}{2008}]%
        {Kuzelka:2009:RSE:1497096.1497102}
\bibfield{author}{\bibinfo{person}{Ondrej Kuzelka} {and} \bibinfo{person}{Filip
  Zelezn{\'y}}.} \bibinfo{year}{2008}\natexlab{}.
\newblock \showarticletitle{A Restarted Strategy for Efficient Subsumption
  Testing}.
\newblock \bibinfo{journal}{{\em Fundam. Inform.\/}}  \bibinfo{volume}{89}
  (\bibinfo{year}{2008}), \bibinfo{pages}{95--109}.
\newblock


\bibitem[\protect\citeauthoryear{Lao, Minkov, and Cohen}{Lao
  et~al\mbox{.}}{2015}]%
        {lao-etal-2015-learning}
\bibfield{author}{\bibinfo{person}{Ni Lao}, \bibinfo{person}{Einat Minkov},
  {and} \bibinfo{person}{William Cohen}.} \bibinfo{year}{2015}\natexlab{}.
\newblock \showarticletitle{Learning Relational Features with Backward Random
  Walks}. In \bibinfo{booktitle}{{\em Proceedings of the 53rd Annual Meeting of
  the Association for Computational Linguistics and the 7th International Joint
  Conference on Natural Language Processing (Volume 1: Long Papers)}}.
  \bibinfo{publisher}{Association for Computational Linguistics},
  \bibinfo{address}{Beijing, China}, \bibinfo{pages}{666--675}.
\newblock
\showDOI{%
\url{https://doi.org/10.3115/v1/P15-1065}}


\bibitem[\protect\citeauthoryear{Li, Chan, and Maier}{Li et~al\mbox{.}}{2015}]%
        {Maier:VLDB:2015}
\bibfield{author}{\bibinfo{person}{Hao Li}, \bibinfo{person}{Chee~Yong Chan},
  {and} \bibinfo{person}{David Maier}.} \bibinfo{year}{2015}\natexlab{}.
\newblock \showarticletitle{Query From Examples: An Iterative, Data-Driven
  Approach to Query Construction}.
\newblock \bibinfo{journal}{{\em PVLDB\/}}  \bibinfo{volume}{8}
  (\bibinfo{year}{2015}), \bibinfo{pages}{2158--2169}.
\newblock


\bibitem[\protect\citeauthoryear{Malec, Khot, Nagy, Blasch, and
  Natarajan}{Malec et~al\mbox{.}}{2016}]%
        {S1}
\bibfield{author}{\bibinfo{person}{Marcin Malec}, \bibinfo{person}{Tushar
  Khot}, \bibinfo{person}{James Nagy}, \bibinfo{person}{Erik Blasch}, {and}
  \bibinfo{person}{Sriraam Natarajan}.} \bibinfo{year}{2016}\natexlab{}.
\newblock \showarticletitle{Inductive logic programming meets relational
  databases: An application to statistical relational learning}. In
  \bibinfo{booktitle}{{\em ILP}}.
\newblock


\bibitem[\protect\citeauthoryear{Mccreath and Sharma}{Mccreath and
  Sharma}{1995}]%
        {Mccreath95extractionof}
\bibfield{author}{\bibinfo{person}{Eric Mccreath} {and} \bibinfo{person}{Arun
  Sharma}.} \bibinfo{year}{1995}\natexlab{}.
\newblock \showarticletitle{Extraction of Meta-Knowledge to Restrict the
  Hypothesis Space for ILP Systems}. In \bibinfo{booktitle}{{\em Australian
  Joint Conference on AI}}.
\newblock


\bibitem[\protect\citeauthoryear{Mitchell}{Mitchell}{1997}]%
        {MachineLearning:Mitchell}
\bibfield{author}{\bibinfo{person}{Tom Mitchell}.}
  \bibinfo{year}{1997}\natexlab{}.
\newblock \bibinfo{booktitle}{{\em Machine Learning}}.
\newblock \bibinfo{publisher}{McGraw-Hil}.
\newblock


\bibitem[\protect\citeauthoryear{Muggleton}{Muggleton}{1995}]%
        {progol}
\bibfield{author}{\bibinfo{person}{Stephen Muggleton}.}
  \bibinfo{year}{1995}\natexlab{}.
\newblock \showarticletitle{Inverse entailment and {Progol}}.
\newblock \bibinfo{journal}{{\em New Generation Computing\/}}
  \bibinfo{volume}{13} (\bibinfo{year}{1995}), \bibinfo{pages}{245--286}.
\newblock


\bibitem[\protect\citeauthoryear{Muggleton and Feng}{Muggleton and
  Feng}{1990}]%
        {golem}
\bibfield{author}{\bibinfo{person}{Stephen Muggleton} {and}
  \bibinfo{person}{Cao Feng}.} \bibinfo{year}{1990}\natexlab{}.
\newblock \showarticletitle{Efficient Induction of Logic Programs}. In
  \bibinfo{booktitle}{{\em ALT}}.
\newblock


\bibitem[\protect\citeauthoryear{Muggleton, Raedt, Poole, Bratko, Flach, Inoue,
  and Srinivasan}{Muggleton et~al\mbox{.}}{2011}]%
        {ILP20:2012}
\bibfield{author}{\bibinfo{person}{Stephen Muggleton}, \bibinfo{person}{Luc~De
  Raedt}, \bibinfo{person}{David Poole}, \bibinfo{person}{Ivan Bratko},
  \bibinfo{person}{Peter~A. Flach}, \bibinfo{person}{Katsumi Inoue}, {and}
  \bibinfo{person}{Ashwin Srinivasan}.} \bibinfo{year}{2011}\natexlab{}.
\newblock \showarticletitle{{ILP} turns 20}.
\newblock \bibinfo{journal}{{\em Machine Learning\/}}  \bibinfo{volume}{86}
  (\bibinfo{year}{2011}), \bibinfo{pages}{3--23}.
\newblock


\bibitem[\protect\citeauthoryear{Muggleton, Santos, and
  Tamaddoni-Nezhad}{Muggleton et~al\mbox{.}}{2009}]%
        {progolem}
\bibfield{author}{\bibinfo{person}{Stephen Muggleton}, \bibinfo{person}{Jose
  Santos}, {and} \bibinfo{person}{Alireza Tamaddoni-Nezhad}.}
  \bibinfo{year}{2009}\natexlab{}.
\newblock \showarticletitle{{ProGolem}: A System Based on Relative Minimal
  Generalisation}. In \bibinfo{booktitle}{{\em ILP}}.
\newblock


\bibitem[\protect\citeauthoryear{Olken}{Olken}{1993}]%
        {olkenThesis}
\bibfield{author}{\bibinfo{person}{Frank Olken}.}
  \bibinfo{year}{1993}\natexlab{}.
\newblock {\em \bibinfo{title}{Random Sampling from Databases}}.
\newblock \bibinfo{thesistype}{Ph.D. Dissertation}. \bibinfo{school}{UC
  Berkeley}.
\newblock


\bibitem[\protect\citeauthoryear{Papenbrock, Kruse, Quian{\'e}-Ruiz, and
  Naumann}{Papenbrock et~al\mbox{.}}{2015}]%
        {Papenbrock:2015:DCI:2752939.2752946}
\bibfield{author}{\bibinfo{person}{Thorsten Papenbrock},
  \bibinfo{person}{Sebastian Kruse}, \bibinfo{person}{Jorge-Arnulfo
  Quian{\'e}-Ruiz}, {and} \bibinfo{person}{Felix Naumann}.}
  \bibinfo{year}{2015}\natexlab{}.
\newblock \showarticletitle{Divide \& Conquer-based Inclusion Dependency
  Discovery}.
\newblock \bibinfo{journal}{{\em PVLDB\/}}  \bibinfo{volume}{8}
  (\bibinfo{year}{2015}), \bibinfo{pages}{774--785}.
\newblock


\bibitem[\protect\citeauthoryear{Picado, Termehchy, and Fern}{Picado
  et~al\mbox{.}}{2017}]%
        {castor:SIGMOD17}
\bibfield{author}{\bibinfo{person}{Jose Picado}, \bibinfo{person}{Arash
  Termehchy}, {and} \bibinfo{person}{Alan Fern}.}
  \bibinfo{year}{2017}\natexlab{}.
\newblock \showarticletitle{Schema Independent Relational Learning}. In
  \bibinfo{booktitle}{{\em SIGMOD Conference}}.
\newblock


\bibitem[\protect\citeauthoryear{Quinlan}{Quinlan}{1990}]%
        {Quinlan:FOIL}
\bibfield{author}{\bibinfo{person}{J.~Ross Quinlan}.}
  \bibinfo{year}{1990}\natexlab{}.
\newblock \showarticletitle{Learning Logical Definitions from Relations}.
\newblock \bibinfo{journal}{{\em Machine Learning\/}}  \bibinfo{volume}{5}
  (\bibinfo{year}{1990}), \bibinfo{pages}{239--266}.
\newblock


\bibitem[\protect\citeauthoryear{Raedt, Poole, Kersting, and Natarajan}{Raedt
  et~al\mbox{.}}{[n. d.]}]%
        {SRLNIPS}
\bibfield{author}{\bibinfo{person}{Luc~De Raedt}, \bibinfo{person}{David
  Poole}, \bibinfo{person}{Kristian Kersting}, {and} \bibinfo{person}{Sriraam
  Natarajan}.} \bibinfo{year}{[n. d.]}\natexlab{}.
\newblock \showarticletitle{Statistical Relational Artificial Intelligence:
  Logic, Probability and Computation}.
\newblock


\bibitem[\protect\citeauthoryear{Richardson and Domingos}{Richardson and
  Domingos}{2006}]%
        {Richardson:2006:MLN}
\bibfield{author}{\bibinfo{person}{Matthew Richardson} {and}
  \bibinfo{person}{Pedro~M. Domingos}.} \bibinfo{year}{2006}\natexlab{}.
\newblock \showarticletitle{Markov logic networks}.
\newblock \bibinfo{journal}{{\em Machine Learning\/}}  \bibinfo{volume}{62}
  (\bibinfo{year}{2006}), \bibinfo{pages}{107--136}.
\newblock


\bibitem[\protect\citeauthoryear{Srinivasan}{Srinivasan}{2004}]%
        {aleph}
\bibfield{author}{\bibinfo{person}{Ashwin Srinivasan}.}
  \bibinfo{year}{2004}\natexlab{}.
\newblock \bibinfo{booktitle}{{\em {The Aleph Manual}}}.
\newblock


\bibitem[\protect\citeauthoryear{Walker, O'Reilly, Kunapuli, Natarajan, Maclin,
  Page, and Shavlik}{Walker et~al\mbox{.}}{2011}]%
        {Walker:2010:AIS:2022735.2022765}
\bibfield{author}{\bibinfo{person}{Trevor Walker}, \bibinfo{person}{Ciaran
  O'Reilly}, \bibinfo{person}{Gautam Kunapuli}, \bibinfo{person}{Sriraam
  Natarajan}, \bibinfo{person}{Richard Maclin}, \bibinfo{person}{David Page},
  {and} \bibinfo{person}{Jude Shavlik}.} \bibinfo{year}{2011}\natexlab{}.
\newblock \showarticletitle{Automating the Ilp Setup Task: Converting User
  Advice About Specific Examples into General Background Knowledge}. In
  \bibinfo{booktitle}{{\em Proceedings of the 20th International Conference on
  Inductive Logic Programming}} {\em (\bibinfo{series}{ILP'10})}.
  \bibinfo{publisher}{Springer-Verlag}, \bibinfo{address}{Berlin, Heidelberg},
  \bibinfo{pages}{253--268}.
\newblock
\showISBNx{978-3-642-21294-9}
\showURL{%
\url{http://dl.acm.org/citation.cfm?id=2022735.2022765}}


\bibitem[\protect\citeauthoryear{Yin, Han, Yang, and Yu}{Yin
  et~al\mbox{.}}{2004}]%
        {join:crossmine}
\bibfield{author}{\bibinfo{person}{Xiaoxin Yin}, \bibinfo{person}{Jiawei Han},
  \bibinfo{person}{Jiong Yang}, {and} \bibinfo{person}{Philip~S. Yu}.}
  \bibinfo{year}{2004}\natexlab{}.
\newblock \showarticletitle{CrossMine: efficient classification across multiple
  database relations}.
\newblock \bibinfo{journal}{{\em ICDE\/}} (\bibinfo{year}{2004}),
  \bibinfo{pages}{399--410}.
\newblock


\bibitem[\protect\citeauthoryear{Zeng, Patel, and Page}{Zeng
  et~al\mbox{.}}{2014}]%
        {QuickFOIL}
\bibfield{author}{\bibinfo{person}{Qiang Zeng}, \bibinfo{person}{Jignesh~M.
  Patel}, {and} \bibinfo{person}{David Page}.} \bibinfo{year}{2014}\natexlab{}.
\newblock \showarticletitle{{QuickFOIL}: Scalable Inductive Logic Programming}.
\newblock \bibinfo{journal}{{\em PVLDB\/}}  \bibinfo{volume}{8}
  (\bibinfo{year}{2014}), \bibinfo{pages}{197--208}.
\newblock


\bibitem[\protect\citeauthoryear{Zhao, Christensen, Li, Hu, and Yi}{Zhao
  et~al\mbox{.}}{2018}]%
        {Zhao2018RandomSO}
\bibfield{author}{\bibinfo{person}{Zhuoyue Zhao}, \bibinfo{person}{Robert
  Christensen}, \bibinfo{person}{Feifei Li}, \bibinfo{person}{Xiao Hu}, {and}
  \bibinfo{person}{Ke Yi}.} \bibinfo{year}{2018}\natexlab{}.
\newblock \showarticletitle{Random Sampling over Joins Revisited}. In
  \bibinfo{booktitle}{{\em SIGMOD}}.
\newblock


\end{thebibliography}
